\documentclass[lettersize,journal]{IEEEtran}
\usepackage{amsmath,amsfonts}
\usepackage{algorithmic}
\usepackage{algorithm}
\usepackage{array}
\usepackage{subcaption}
\usepackage{textcomp}
\usepackage{stfloats}
\usepackage{url}
\usepackage{verbatim}
\usepackage{array}
\usepackage{booktabs}
\usepackage{siunitx}

\usepackage{cite}
\usepackage{float}
\usepackage{todonotes}
\usepackage{comment}
\usepackage{mathtools}
\usepackage[pdfencoding=auto,hidelinks]{hyperref}

\begin{document}

\title{Characterizing the Sensitivity to Individual Bit Flips in Client-Side Operations of the CKKS Scheme}

\author{
    \IEEEauthorblockN{Mat\'ias Mazzanti\textsuperscript{1,2}, Augusto Vega\textsuperscript{3}, Pradip Bose\textsuperscript{3},  Esteban Mocskos\textsuperscript{1,2}
    \\}
    \IEEEauthorblockA{
        \textit{\textsuperscript{1}Departamento de Computaci\'on, Facultad de Ciencias Exactas y Naturales, Universidad de Buenos Aires (Argentina), \textsuperscript{2}Centro de Simulaci\'on Computacional p/Aplic Tecnol\'ogicas (CSC-CONCICET), \textsuperscript{3}IBM T. J. Watson Research Center (NY, USA)}}
}

\markboth{}%
{Mazzanti \MakeLowercase{\textit{et al.}}: Characterizing the Sensitivity to Individual Bit Flips in Client-Side Operations of the CKKS Scheme}


\maketitle

\begin{abstract}
Homomorphic Encryption (HE) enables computation on encrypted data without decryption, making it a cornerstone of privacy-preserving computation in untrusted environments.
As HE sees growing adoption in sensitive applications--such as secure machine learning and confidential data analysis--ensuring its robustness against errors becomes critical.
Faults (e.g., transmission errors, hardware malfunctions, or synchronization failures) can corrupt encrypted data and compromise the integrity of HE operations.

Among these, the specific impact of soft errors (such as bit-flips) on modern HE schemes has received little attention.
Specifically, the CKKS scheme--one of the most widely used HE schemes for approximate arithmetic--lacks a systematic study of how such errors propagate across its pipeline, particularly under optimizations like the Residue Number System (RNS) and Number Theoretic Transform (NTT).

This work bridges this gap by presenting a theoretical and empirical analysis of CKKS’s fault tolerance behavior under bit-flip errors. We focus on client-side operations (encoding, encryption, decryption, and decoding) and demonstrate that while the vanilla CKKS scheme exhibits some resilience, performance optimizations like RNS and NTT exacerbate the error sensitivity of the scheme. By characterizing these failure modes, we lay the groundwork for developing HE hardware and software systems that are resilient to errors, ensuring both performance and integrity in privacy-critical applications.
\end{abstract}


\section{Introduction}
\label{section:introduction}

\IEEEPARstart{T}{he} need to preserve data privacy becomes critical when processing occurs outside the client's trusted environment, such as in public clouds or on third-party platforms.
In these scenarios, protecting data not only at rest or in transit, but also during processing, is a key challenge. Homomorphic Encryption (HE) emerges as a promising technique to address this issue by enabling \textit{computation on encrypted data} (i.e., without requiring decryption), thereby ensuring confidentiality at \textit{all} times.
For example, it allows a cloud server to calculate statistics on encrypted medical records or train a machine learning model using financial data without revealing its content. In these cases, the server performs the required operations without accessing the plaintext data, preserving its confidentiality throughout the process.

This ability to operate directly on encrypted data is not merely convenient; it is a fundamental requirement for enabling secure computation in infrastructures where full trust cannot be assumed, such as public clouds, third-party platforms, or shared devices. Without such schemes, sensitive applications requiring delegated processing would be infeasible from a privacy standpoint. Among the most widely used homomorphic schemes is CKKS (Cheon-Kim-Kim-Song)~\cite{Cheon2017} due to its fixed-point arithmetic support. This makes it ideal to handle real number computation in tasks such as numerical analysis, signal processing, or machine learning.
\begin{figure}[!hb]
  \centering
  \includegraphics[width=0.9\linewidth]{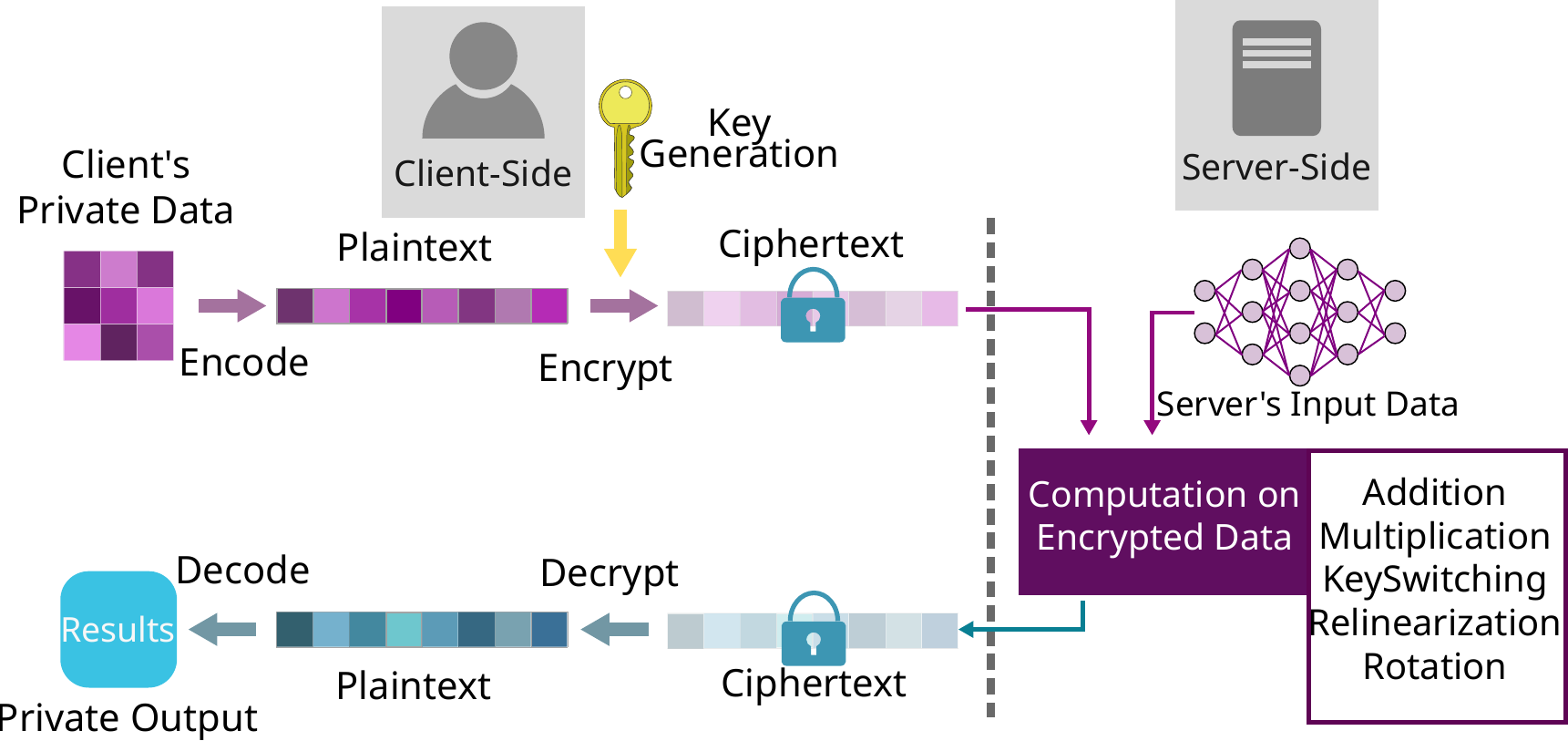}
  \caption[Typical HE use case]{Typical HE use case: a client sends encrypted data to the cloud,
    the cloud operates on the encrypted data and returns the results.
    Data and results remain encrypted all the time.\footnotemark}
  \label{fig:cloudFHE}
\end{figure}
\footnotetext{Adapted from~\url{https://openmined.org/blog/from-fully-homomorphic-encryption-to-silicon/}}
An encryption scheme should support encrypted computation of at least addition and multiplication operations in order to be considered \textit{homomorphic}. These operations are sufficient to construct any arithmetic circuit, thereby allowing arbitrary computation without decrypting the information.
In addition to these fundamental operations, some modern schemes incorporate additional primitives--such as elements rotation or complex conjugation--which, while not increasing the scheme's expressiveness, are crucial for improving efficiency and facilitating certain computational structures, especially in applications like machine learning. HE schemes can be divided into three main categories:
\begin{itemize}

\item \textit{Partial Homomorphic Encryption} (PHE) supports a single type of operation on encrypted data, either addition or multiplication. An example is The RSA (Rivest–Shamir–Adleman) cryptosystem\cite{Rivest1978}, which supports multiplication of encrypted data but not addition.

\item \textit{Leveled Homomorphic Encryption} (LHE) supports both operations (addition and multiplication), but only for a limited number of steps before the accumulated noise makes correct decryption infeasible.

\item \textit{Fully Homomorphic Encryption} (FHE) enables unlimited additions and multiplications, enabling arbitrary computation on encrypted data.
\end{itemize}

As illustrated in Figure~\ref{fig:cloudFHE}, the typical process for an HE scheme begins with the client encrypting data locally using a secret key before sending it to the cloud server.
The server processes this data homomorphically; that is, it performs computations directly on the encrypted data. Once processing is complete, the result, which remains encrypted, is returned to the client, who is solely responsible for its decryption.

The CKKS scheme is gaining increasing interest in machine learning and data analysis applications due to its ability to perform approximate calculations on floating-point numbers.
This characteristic is fundamental for efficiently handling large data volumes and training predictive models, allowing sensitive data to be processed securely without needing decryption.
Despite its high computational cost, its potential in these fields is still an active area of research and development~\cite{Ma2022,Zhang2023,Zhang2024}.

One of the most significant enhancements to the CKKS scheme is using the Residue Number System (RNS) alongside the Number Theoretic Transform (NTT). This combination has helped establish it as a standard and leads to what is known as the \textit{full} RNS variant~\cite{Cheon2019}.
Throughout this paper, we will refer to this variant simply as CKKS, while the original version without these optimizations will be termed \emph{vanilla CKKS}.

\begin{figure}[!ht]
  \centering
  \begin{subfigure}[!ht]{0.49\linewidth}
    \centering
    \includegraphics[height=2cm]{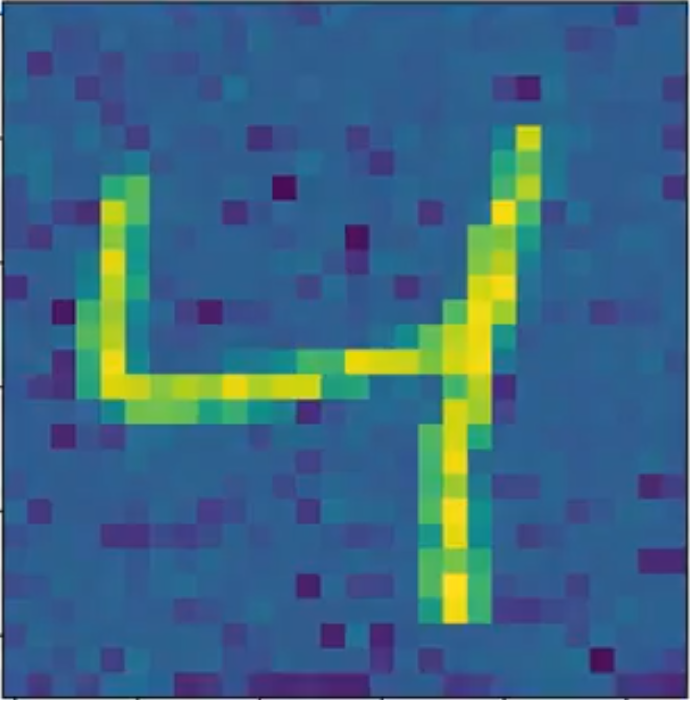}  
    \caption{Original}
    \label{fig:decryptable}
  \end{subfigure}
  \begin{subfigure}[!ht]{0.49\linewidth}
    \centering
    \includegraphics[height=2cm]{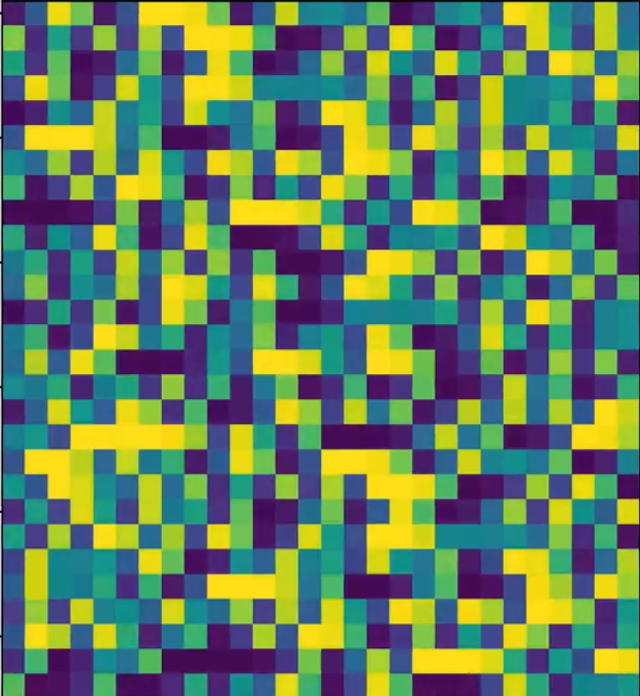}  
    \caption{RNS and NTT enabled}
    \label{fig:undecryptable}
  \end{subfigure}
  \caption{Example of the impact of a single bit-flip in CKKS.}
  \label{fig:ckks-error-example}
\end{figure}

It is crucial to recognize that the CKKS scheme is vulnerable to errors caused by single bit-flips, which can compromise both computational accuracy and data integrity.
Since bit-flips may occur under various conditions, assessing their impact on CKKS-based applications--particularly in unreliable hardware environments--is crucial.
Figure~\ref{fig:ckks-error-example} illustrates a scenario in which an image from the MNIST dataset (Figure~\ref{fig:decryptable}) is processed through the CKKS pipeline.
Figure~\ref{fig:undecryptable} shows an entirely distorted image recovered when a single bit error is introduced during the encoding phase using RNS and NTT.
These errors can originate in hardware due to phenomena such as interference, memory failures, or even fault injection attacks~\cite{Sabath2010,Kolditz2014,Mutlu2019rowhammer,Li2020,Li2020b,Hoeffgen2020,Dixit2021}, making bit-flip sensitivity analysis a critical aspect for the practical and reliable implementation of HE.

In this work, we conduct a comprehensive analysis of the CKKS scheme's sensitivity to single bit errors in client-side stages, such as encoding, encryption, decryption, and decoding operations. We focus on the impact of RNS (Residue Number System) and NTT (Number Theoretic Transform) optimizations on the scheme's resilience against these errors. To the best of our knowledge, this paper is the first study that explores this aspect of HE.
Our analysis provides a general framework aimed at enhancing the security and robustness of CKKS in diverse contexts. Our main contributions are as follows:

\begin{itemize}
\item We present the first comprehensive analysis of the CKKS scheme's sensitivity to individual bit-flip errors in a typical client-side pipeline, considering configurations with and without RNS and NTT optimizations.

\item We evaluate the impact of RNS and NTT optimizations on the robustness of the CKKS scheme.

\item We identify specific configurations where the use of certain optimizations introduces a form of intrinsic redundancy, increasing the proportion of bits that remain correct in the presence of errors.
This finding is particularly relevant for applications where fault tolerance is a fundamental requirement.

\item We provide a framework that can serve as a foundation for future research aimed at improving the security and resilience of CKKS against hardware failures.
\end{itemize}

Section~\ref{section:background} describes the CKKS scheme in its vanilla and full variants. Section~\ref{section:error-resilience-analysis} presents a theoretical analysis of CKKS's robustness against bit-flip errors.
Section~\ref{section:methodology} details the methodology used in our experimental campaigns.
Section~\ref{section:results} develops the experimental analysis and discusses our findings. Finally, Section~\ref{section:conclusions} presents our conclusions.

\section{Background and CKKS Overview}\label{section:background}

In this section, we first introduce the core parameters of the CKKS scheme, followed by a conceptual overview alongside a simplified explanation of its key components--making the material accessible even to readers without a specialized background in cryptography. We then delve into the mathematical foundations required for the analysis presented in Section~\ref{section:error-resilience-analysis}.

\subsection{General Flow of the CKKS Scheme}\label{sec:ckks-overview}

CKKS is an approximate homomorphic encryption scheme designed to enable computations on encrypted data while providing explicit control over approximation error. It operates over a polynomial ring of degree $N$ based on the Ring Learning With Errors (RLWE) problem, a lattice-based problem that ensures security through the controlled injection of noise into ciphertexts.
Figure~\ref{fig:cloudFHE} shows the processing stages:


\begin{enumerate}
\item \textbf{Key Generation:} The secret key and public key are generated as degree-$N$ polynomials, created through random sampling and error injection.
\item \textbf{Encoding:} The input vector is transformed using a specialized inverse Fast Fourier Transform (IFFT) into a degree-$N$ polynomial, constructed such that its evaluations at certain roots of unity approximately reproduce the original values.
The resulting polynomial's coefficients are then scaled by $\Delta$ and rounded to integers, producing the \emph{plaintext}.
Formally, given a polynomial $P$ and a vector $z \in \mathbb{R}^{N}$, the following relation holds:
\begin{equation*}
\boldsymbol{z} \approx (P(\xi), P(\xi^3), \cdots, P(\xi^{2N-1})).
\end{equation*}

A specialized version of the FFT, adapted to the ring $\mathbb{Z}_q[X]/(X^N+1)$, is used to preserve the algebraic structure of the scheme and enable various optimizations.

\item \textbf{Encryption:} Given the plaintext, two encrypted polynomials $(c_0, c_1)$ are generated using the public (or secret) key and by adding a small amount of noise.
The result $c = (c_0, c_1)$ is referred to as the \emph{ciphertext}.

\item \textbf{Computation:} Homomorphic operations are performed directly on the ciphertexts, including addition, multiplication, rotations, and other transformations compatible with the scheme.

\item \textbf{Decryption and Decoding:} One component of the ciphertext is multiplied by the secret key and added to the other component.
The result is a plaintext, which is divided by $\Delta$, and then a specialized FFT is applied to recover an approximation of the original input vector.
\end{enumerate}

Since the security of CKKS relies on the computational hardness of the RLWE problem, each encryption inherently introduces noise.
Although it is small at the beginning, it accumulates during homomorphic operations.
Each operation increases the noise level in the ciphertext, with homomorphic multiplication contributing the most to this growth. For this reason, the number of successive multiplications is often used as an indirect estimate of the accumulated error in a ciphertext.
If this growth is not properly managed--e.g., through techniques such as \textit{bootstrapping}~\cite{Cheon2018}--the error can exceed tolerable limits, resulting in incorrect outputs during decryption and decoding.

\subsection{Parameters controlling CKKS}
\label{sec:ckks-parameters}

CKKS operates over a polynomial ring defined as:
\begin{equation*}
\mathcal{R}_q = \mathbb{Z}_q[X]/(X^N + 1),
\end{equation*}
where \(\mathbb{Z}_q\) denotes the integers modulo \(q\) (i.e., the integer values between \(0\) and \(q - 1\), with addition and multiplication performed modulo \(q\))); \(\mathbb{Z}_q[X]\) denotes the polynomials with coefficients in \(\mathbb{Z}_q\); and \(/(X^N + 1)\) indicates that polynomials are considered modulo \(X^N + 1\).

This ring imposes a negacyclic structure that enables, after encoding, the efficient representation of vectors of complex numbers.
In CKKS, data is encoded as a polynomial that, when evaluated at specific values, yields each of the encoded data points. In particular, the degree $N$ must be a power of two and serves two critical purposes: it determines the maximum number of complex values that can be encoded simultaneously, and enforces the security of the scheme.

The original (vanilla) version of the CKKS scheme uses the following core parameters, which are selected to achieve the desired security level, maintain tolerable error growth, and minimize computational cost:
\begin{itemize}
\item $N$ is the ring degree, typically in the range $2^{12}-2^{16}$, and it determines both the dimension of the encoding space and the computational cost.

\item $q_0$ is the initial modulus--a single large integer (typically up to 60 bits)--that determines the target precision.

\item $\Delta$ is the scaling factor, a constant used to amplify real values before encoding to preserve floating-point precision.

\item $L$ is the multiplicative depth; i.e., the number of homomorphic multiplications that can be performed before accumulated noise degrades precision.
\end{itemize}

The total modulus of the scheme, denoted $Q$, and the size of the coefficients in the encoding and encryption polynomials are defined differently depending on the CKKS variant used.
In the vanilla version, $Q = \Delta^L \cdot q_0$.

In CKKS, each multiplication between ciphertexts increases both the error and the magnitude of the encoded values.
To control this growth and preserve precision, an operation called \emph{rescaling} is applied.
Rescaling reduces the scale of the encoded values and also decreases the active modulus, which in turn limits the number of consecutive multiplications (with rescaling) that can be performed.
This limit is known as the \emph{multiplicative depth} of the scheme.
The factor $\Delta$ appears once at each of the $L$ multiplicative levels of the modulus hierarchy, which explains the term $\Delta^L$ in the expression for $Q$.

\subsection{Full CKKS Variant}
\label{sec:full-ckks-variant}

The magnitude of the modulus $Q$ (and consequently, the size of the polynomial coefficients) can range from hundreds to thousands of bits, depending on the use case. However, modern computing systems typically operate with word sizes of 64 bits or less, making direct operations on such large coefficients impractical. The \textit{Residue Number System} (RNS) enables the representation of integer values using their remainders with respect to a set of pairwise coprime moduli. This allows arithmetic operations--such as addition and multiplication--to be performed using 64-bit word size, reducing computational complexity and significantly improving efficiency.
These properties make RNS particularly well-suited for high-performance contexts such as homomorphic encryption. The use of the \textit{Chinese Remainder Theorem} (CRT) is essential in RNS, as it ensures that each set of residues uniquely corresponds to a single integer within a defined range. Moreover, CRT provides an efficient procedure to reconstruct the original integer from its residue representation.

Once mapped to the Residue Number System (RNS), each polynomial is further transformed via the Number Theoretic Transform (NTT)--the modular analogue of the Fast Fourier Transform (FFT). This reduces the complexity of polynomial multiplication (one of the most computationally demanding operations in the scheme) from $\mathcal{O}(n^2)$ to $\mathcal{O}(n\log n)$.
By default, the CKKS scheme assumes that polynomials are represented in the evaluation domain (NTT). When certain operations require polynomials to be in the coefficient domain, an inverse NTT (iNTT) is applied to transform the polynomial from the evaluation domain back to the coefficient domain.

These optimizations (RNS and NTT) do not alter the semantics of the encryption scheme but do change its internal implementation.
In particular, they have implications for error propagation and sensitivity, topics that will be analyzed in detail in Section~\ref{section:error-resilience-analysis}.

In the full CKKS variant, the total modulus of the scheme is expressed as:
\begin{equation}\label{eq:QRNS}
Q = \prod_{i=0}^{L-1} q_i,
\end{equation}
where $q_0$ is the initial modulus and $q_1, \dots, q_{L-1}$ are additional moduli that in general are selected by the scheme, all of similar bit length.
It is worth noting that some libraries provide manual control over the bit-length of these $q_i$ moduli; the implementations employed in this work automatically configure these parameters according to the bit length specified for $q_0$.

\subsection{Encoding/Encryption/Decryption/Decoding}
\label{subsec:insights}

The CKKS scheme operates on complex vectors of size up to $N/2$. This input is encoded into the polynomial ring $\mathbb{Z}[X]/(X^N + 1)$, producing a plaintext polynomial $m$.
The resulting plaintext is then encrypted using either the public or secret key, following the RLWE problem, where a small amount of noise is intentionally injected for security.
The ring $\mathbb{Z}[X]/(X^N + 1)$ is preferred in RLWE-based schemes due to its stronger security properties compared to $\mathbb{Z}[X]/(X^N - 1)$, where the convolution is cyclic and considered more vulnerable to attacks.

\paragraph{Precision}

To better understand the parameter selection in CKKS, it is helpful to clarify the concept of precision.
Some CKKS implementations use real-valued inputs only, the scaling parameter $\Delta$ determines how many bits are used to encode the fractional part of the input values, and the difference between the initial modulus $q_0$ and $\Delta$ indicates the number of bits allocated to the integer part. For example, using $q_0 = 60$ bits and $\Delta = 50$ bits allows the encrypted representation of real numbers with approximately 10 bits of precision for the integer part and 50 bits for the fractional part.

\paragraph{Keys}
Encryption in the CKKS scheme requires either a secret key $sk$ or a public key $pk$.
The secret key is generated as $sk=(1,s)$ by sampling a vector $s$ of length $N$ generally from a ternary distribution with Hamming weight $h$. The public key is constructed by sampling $a_0$ from $\mathbb{Z}_Q$ (i.e., from the set of positive integers less than the modulus $Q$) and an error term $e_0$ from a discrete Gaussian distribution with standard deviation $\sigma = 3.2$, as follows:
\begin{equation*}
pk = ([-a_0s + e_0]_Q, a_0) = (p_0, p_1)
\end{equation*}
where the notation $[\cdots]_Q$ denotes reduction modulo $Q$.

\paragraph{Encoding}

The key idea behind encoding in CKKS is to transform the input vector into a polynomial.
Different libraries implementing CKKS vary in how they perform this transformation using IFFT/FFT-based methods. In all cases, the standard cyclic transform must be adapted to a negacyclic version to ensure that operations remain within the ring $\mathbb{Z}_q[X]/(X^N + 1)$ used by the scheme.

The transforms employed by the most adopted CKKS implentations are negacyclic and include scheme-specific optimizations, commonly referred to as specialized FFT/IFFT.
These versions incorporate several improvements originally proposed by Bernstein~\cite{Bernstein2007,Bernstein2008}.
One of their main advantages is that they allow the transform to be applied directly to vectors of size $\leq N/2$, without expanding them to size $N$, which significantly reduces computational cost.

Given a complex-valued input vector of size $n \leq N/2$, the encoding process begins by padding the vector with zeros to reach the next power of two, denoted $n'$.
A specialized inverse FFT is then applied, producing a complex vector of length $n'$.
From this, a polynomial of degree $N-1$ is constructed.

\begin{figure}[!ht]
  \centering
  \includegraphics[scale=0.28]{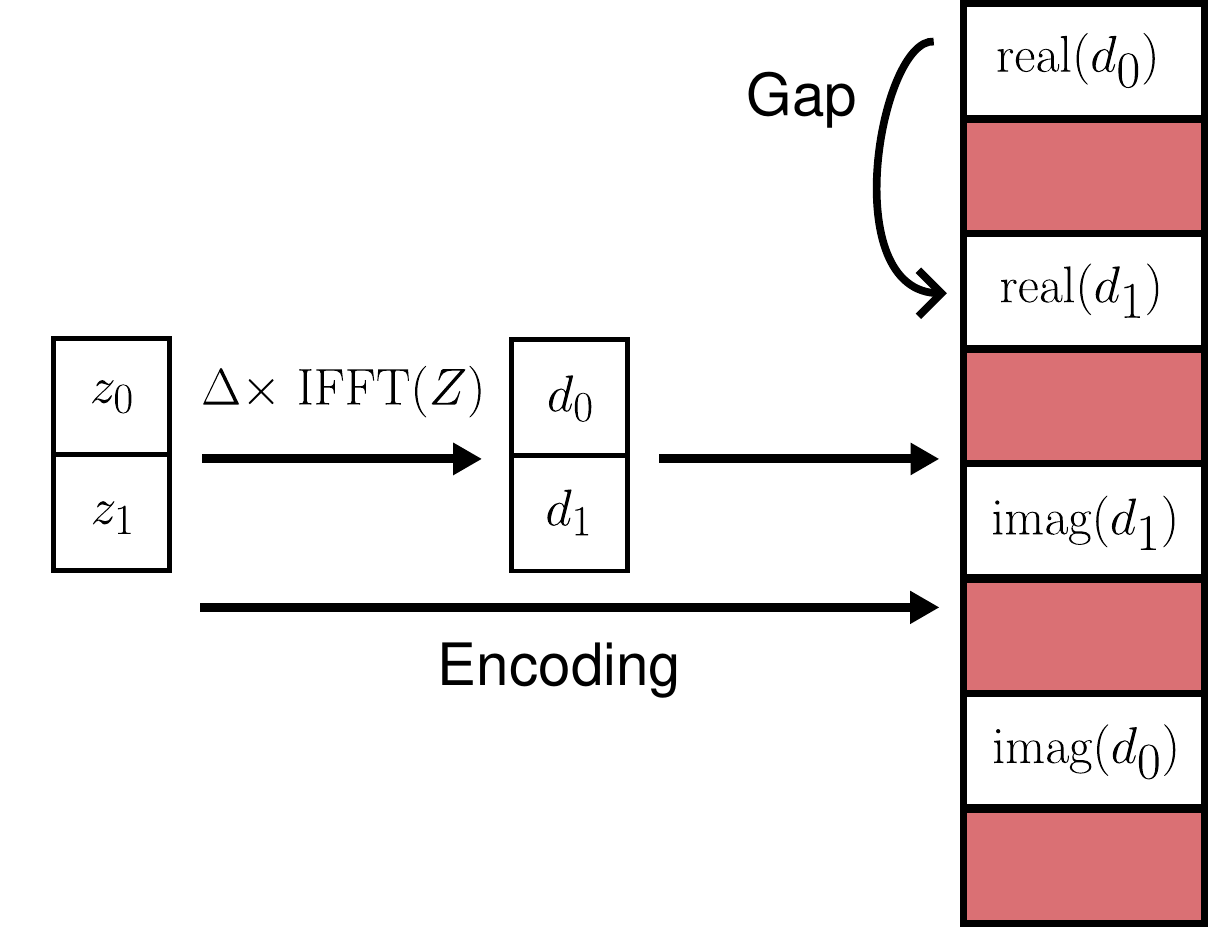}
  \caption{Standard encoding scheme in CKKS. The example shows an input vector with $n = 2$ values, represented in a polynomial ring of size $N=8$, where the gap between consecutive coefficients is 2.}
  \label{fig:encodingScheme}
\end{figure}

First, the gap parameter $gap = \frac{N/2}{n'}$ is computed, and the lower half of the polynomial (coefficients of degree 0 to $N/2 - 1$) is filled with the real parts of the transformed vector, with assignments spaced at intervals of size \textit{gap}. The upper half does the same using the imaginary part of the transformed vector.
Unassigned coefficients are set to zero. This process is depicted in Figure~\ref{fig:encodingScheme}. As part of this encoding stage, each coefficient is multiplied by $\Delta$, and then rounded to the nearest integer using the coordinate-wise randomized rounding method described in~\cite{Lyubashevsky2013b}.
\paragraph{Encryption}

The encryption of a plaintext $m$ using the secret key $sk=(1,s)$ and modulus $Q$ is given by:
\begin{equation}\label{eq:cipherSecret}
    c_{sk} = ([m + a\times s + e]_Q, [-a]_Q) = (c_0, c_1)
\end{equation}
where $a$ is a polynomial sampled from a uniform distribution, $e$ is a noise term sampled from a discrete Gaussian distribution, and the notation $[\cdots]_Q$ denotes reduction modulo $Q$.

An alternative procedure is to encrypt by using the public key, $pk=(p_0, p_1)$.
In this case, encryption is defined as:
\begin{equation}\label{eq:cipherPublic}
    c_{pk} = ([m + p_0\times v + e_1]_Q, [p_1\times v + e_2]_Q) = (c_0, c_1)
\end{equation}
where $v$ is an ephemeral (one-time) polynomial sampled from a ternary distribution, and $e_1$, $e_2$ are error terms sampled similarly to $e$ in the previous procedure.

\paragraph{Decryption}

Decryption in both secret key and public key cases uses the secret key $sk=(1,s)$ and the encrypted text $(c_0, c_1)$ as follows:
\begin{equation}
    m = [c_0 + c_1\times s]_Q \label{eq:desencrypt}
\end{equation}

\paragraph{Decoding}

Decoding a plaintext polynomial $m$ with $N$ coefficients implies reconstructing a complex-valued vector of $n'$ elements by reversing the encoding process illustrated in Figure~\ref{fig:encodingScheme}.
The $i$-th element ($i\leq N/2$) of the resulting vector is formed by using the real part of the $i$-th coefficient of the plaintext, and the imaginary part from the coefficient at position $i+N/2$. This complex vector is then divided by $\Delta$, and a specialized FFT is applied to recover the approximate original input.

Even when simply encoding and decoding an input vector without performing any computation, small approximation errors may arise due to rounding effects.
Corollary 3 in~\cite{Costache2024} shows that the worst-case distance between the input vector $z$ and the output vector $z'$ (after encoding and decoding) at each position is approximately: $N / (\pi \times \Delta)$. With a sufficiently large $\Delta$, this difference becomes negligible.
Given that a typical ring size $N$ ranges from $2^{12}-2^{16}$, using a scaling factor on the order of $2^{20}$ is enough to ensure a negligible approximation error.



\section{Error Resilience Analysis} \label{section:error-resilience-analysis}

This section presents a theoretical analysis of the error induced by a single bit flip at different stages of the CKKS scheme, in its vanilla and full RNS variants. The goal of this analysis is:
\begin{itemize}
\item Determining analytical error bounds in the vanilla variant (no RNS/NTT).
\item Identifying the most sensitive polynomial coefficients.
\item Enabling a better understanding of error behavior in practical implementations.
\end{itemize}

Although a closed-form analytical bound cannot be derived for the full RNS variant due to the complexity of modular representations, a qualitative analysis of the observed patterns is provided.

\subsection{Error Model for a Single Bit Flip}
\label{subsec:teosingleBitFlip}

We adopt a single bit-flip (soft error) model, where a single bit is altered. In practice, such faults may be caused by electromagnetic interference~\cite{Sabath2010,Li2020}, hardware failures~\cite{VanderLeest2010,Kolditz2014}, or cosmic  rays~\cite{Ziegler1979,Hoeffgen2020}.

Let $p$ be the original value of a coefficient, either in a plaintext polynomial $\boldsymbol{m}$ or in one of the ciphertext polynomials $(c_0,c_1)$.
We denote this generic polynomial as $\mathbf{p}$.
Let's $p_j'$ be the result of flipping a bit $j$ of $p$, then the resulting error is:
\begin{equation}
e_j = p_j' - p
\end{equation}

We model this fault as an error polynomial $\boldsymbol{e}_{i,j}\in\mathbb{Z}[X]$, whose only non-zero coefficient is the $(i-1)$-th, corresponding to an error at bit $j$ of coefficient $i-1$:
\begin{equation}
\boldsymbol{e}_{i,j} = e_j\,X^{i-1}
\end{equation}

This model enables us to track how a single bit flip propagates through the cryptographic pipeline.

\subsection{CKKS Encoding via DFT} \label{subsec:teoerrorPlaintex}

As discussed in Section~\ref{section:background}, the Fast Fourier Transform (FFT) is one of the key components of encoding and decoding in CKKS. Although CKKS uses a specialized inverse FFT for computational efficiency, this transformation is typically implemented using optimized algorithms that obscure its underlying linear structure. Instead, we rely on the negacyclic Discrete Fourier Transform (DFT), an algebraically equivalent transformation explicitly expressed as matrix multiplication.
This formulation allows us to:
\begin{enumerate}
\item Directly represent how errors propagate through the encoding process.
\item Analyze the isolated effect of a perturbation in a single coefficient.
\item Derive analytical bounds on the magnitude of the error without dealing with the internal optimizations of FFT implementations.
\end{enumerate}

While this choice sacrifices performance, it offers significant pedagogical advantages, allowing for an intuitive understanding of how bit-flip errors are amplified or distributed during the encoding stage.

\subsubsection{Encoding via Negacyclic DFT} \label{subsec:teoerrorDFT}

Let $M=2N$ and $\xi = e^{-2\pi i / M}$.
To construct a negacyclic Discrete Fourier Transform (DFT), we evaluate it at the odd powers of the roots of the cyclotomic polynomial $\Phi_M(X)=X^N + 1$; that is, at $\xi^{1}, \xi^{3}, \xi^{5}, \dots, \xi^{2N-1}$.
We define the \emph{Vandermonde} matrix where each column corresponds to a different power of these roots of unity:
\begin{equation}
W \;=\;
\begin{pmatrix}
\xi^{1\cdot 0} & \xi^{1\cdot 1} & \cdots & \xi^{1\cdot (N-1)} \\[4pt]
\xi^{3\cdot 0} & \xi^{3\cdot 1} & \cdots & \xi^{3\cdot (N-1)} \\[2pt]
\vdots         & \vdots         & \ddots & \vdots               \\[2pt]
\xi^{(2N-1)\cdot 0} & \xi^{(2N-1)\cdot 1} & \cdots & \xi^{(2N-1)\cdot (N-1)}
\end{pmatrix}
\end{equation}
and we denote its inverse as $W^{-1}$, which is required for the encoding process.
Due to the properties of this matrix, its inverse is simply the conjugate transpose (Hermitian transpose) normalized by $N$:
\begin{equation*}
    W^{-1} = \frac{1}{N}W^{\dagger}
\end{equation*}

To use the DFT and its inverse for encoding to obtain results equivalent to those produced by standard implementations, some preprocessing is required: the input vector must be resized to length $N$ and reordered.
However, these adjustments do not affect the behavior of the error model under analysis and are omitted for simplicity.

In general, for an input vector of $n\leq N/2$ elements, it is first zero-padded to length $N/2$, and then mirrored with its complex conjugate to reach a total length of $N$.
Once the input vector $\boldsymbol{z}\in\mathbb{R}^N$ is prepared and reordered, the encoding proceeds as follows:

\begin{equation}
    \begin{split}
       \mathrm{Encode}(\boldsymbol{z}) &= \mathrm{Round}\bigl(W^{-1}(\Delta \boldsymbol{z})\bigr)\\
        &=  \mathrm{Round}\bigl(\Delta \,\cdot\, W^{-1}\boldsymbol{z}\bigr) = \boldsymbol{m} \in \mathbb{Z}^N,
    \end{split}
\end{equation}
where $\Delta$ is the already defined scaling factor.
Due to the Hermitian symmetry of $W$ and $W^{-1}$, and the conjugate extension of the input, the result of $W^{-1}\boldsymbol{z}$ lies entirely in the real subspace.

To recover an approximation of the original input $z$ from $m$, decoding is performed as:
\begin{equation}
\label{eq:decodeError}
  \mathrm{Decode}(\boldsymbol{m})
  = \boldsymbol{z}'
  = W \,\bigl(\boldsymbol{m}/\Delta\bigr)
\end{equation}
Finally, the output undergoes the same reordering process as in encoding, and the $n$ components of $z'$ are extracted.
Since the DFT output is complex-valued, the real part is extracted to obtain the final approximation of z

\subsection{Theoretical Estimation of Encoding Error on vanilla CKKS}\label{subsec:teoencdec}

This section focuses on the impact of single-bit errors during the encoding phase of the CKKS scheme.
Let us consider the case in which an error $\boldsymbol{e}_{i,j}$ is introduced by flipping a single bit in the coefficient $i$ of the plaintext vector $\boldsymbol{m}$, specifically at the $j$-th bit position.
The resulting perturbed plaintext is $\boldsymbol{m}' = \boldsymbol{m} + \boldsymbol{e}_{i,j}$, and decoding proceeds as in Equation~\ref{eq:decodeError}:
\begin{equation*}
\begin{split}
        \mathbf{z''} &= W \times \frac{\mathbf{m'}}{\Delta} = W \times \frac{\mathbf{m}+ \mathbf{e}_{i,j}}{\Delta}\\
        &= W \times \frac{\mathbf{m}}{\Delta}  + W \times \frac{\mathbf{e}_{i,j}}{\Delta}  \\
        &= \mathbf{z'} + W \times \frac{\mathbf{e}_{i,j}}{\Delta}
\end{split}
\end{equation*}

Since a single bit flip affects only one coefficient, the resulting error is localized.
The difference between the decoded vectors $\boldsymbol{z'}$ and $\boldsymbol{z''}$ is given by:
\begin{equation}\label{eq:analytical_error}
    \lVert \boldsymbol{z'} - \boldsymbol{z''} \rVert \propto
    \lVert W \times \boldsymbol{e}_{i,j} / \Delta \rVert
\end{equation}

\subsubsection{Bit-Flip Injection during Encoding} \label{subsec:teoPlain}

To better understand the analytical expression of the error in Equation~\ref{eq:analytical_error}, we simulate flipping each bit in the plaintext coefficients.
Let $j\in\{0,\dots,63\}$ denotes the bit position in a 64-bit unsigned integer.
For each iteration, we flip the $j$-th bit of a coefficient, introducing an error with integer magnitude $e_j = 2^j$.
This represents the maximum jump in value induced by flipping bit $j$ in a coefficient, which is stored as an unsigned integer (which is the case of plaintext coefficients).

We repeat this process for all coefficients, grouping the results into $N=4$ blocks of 64 iterations each, and using $\Delta = 2^{50}$.
Let $\boldsymbol{z}'$ be the decoded vector without errors, and $\boldsymbol{z}''$ the result after applying the bit flip and decoding.
Using Equation~\ref{eq:analytical_error}, we express the error using the $L_{2}$ norm:
\begin{equation}\label{eq:norm2}
    L_2(i,j) = \bigl\lVert Real(W \times \boldsymbol{e}_{i,j} / \Delta) \bigr\rVert_2
\end{equation}
where $\mathbf{e}_{i,j}$ is a zero vector of length $N$ with value $e_j$ at position $i$.
We also take the real part of the vector before computing the $L_{2}$ norm, to simulate the encoding of real-valued data.
This step is important not only to maintain consistency with the backend used in later sections, but also because it exposes a specific behavior that would otherwise not manifest when keeping the complex part.
\begin{figure}[!ht]
    \centering
    \includegraphics[width=0.9\linewidth]{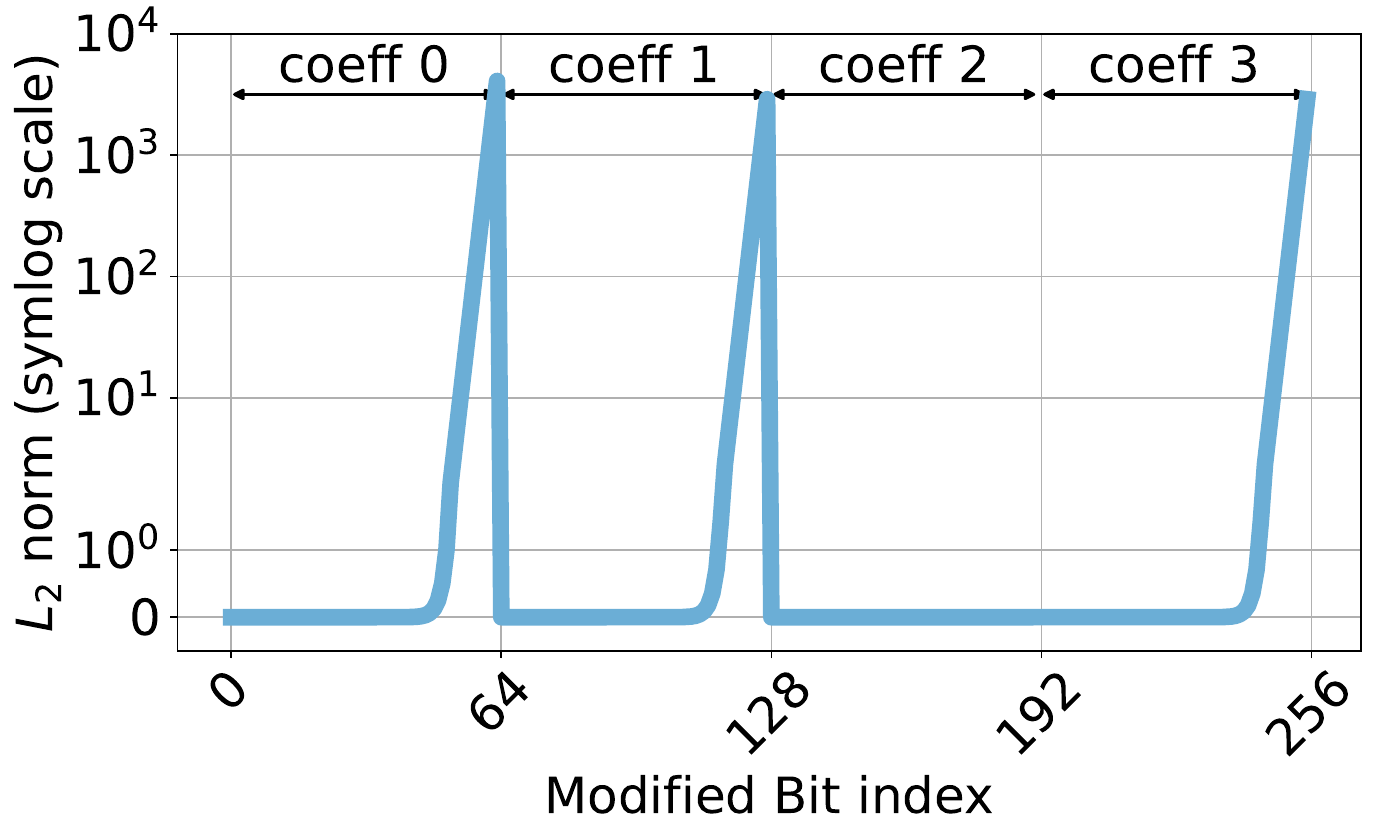}
    \caption{$L_2$-norm of the error, measured as the difference between the golden (no bit-flip) output and the output obtained with a single bit-flip in the plaintext during encoding.
    The horizontal axis shows the bit-flip position. Parameters: $N=4$ and $\Delta = 2^{50}$.
    }
    \label{fig:theoreticNorm}
\end{figure}
Figure~\ref{fig:theoreticNorm} shows the values of $L_2(i,j)$ as a function of the bit-flip position. $L2(i,j)$ grows exponentially with $j$ for each coefficient $i$, consistent with the $2^j$ dependency. An exception appears at  $i = N/2 = 2$, where the norm remains zero. This occurs because, in the DFT, the coefficient at index $N/2$ contributes only imaginary components, which vanish when taking the real part of the transform.

\subsubsection{Effect of the Scaling Factor on Bit-Flip Error} \label{subsec:teoDelta}

To explore how the scaling factor $\Delta$ affects the magnitude of bit-flip errors, we repeat the previous experiment using different values of $\Delta\in\{2^{20},2^{40},2^{50}\}$.
For each case, we plot $L_2(i,j)$ using a semi-logarithmic scale on the $y$-axis for easier comparison.
\begin{figure}[!ht]
    \centering
    \includegraphics[width=0.9\linewidth]{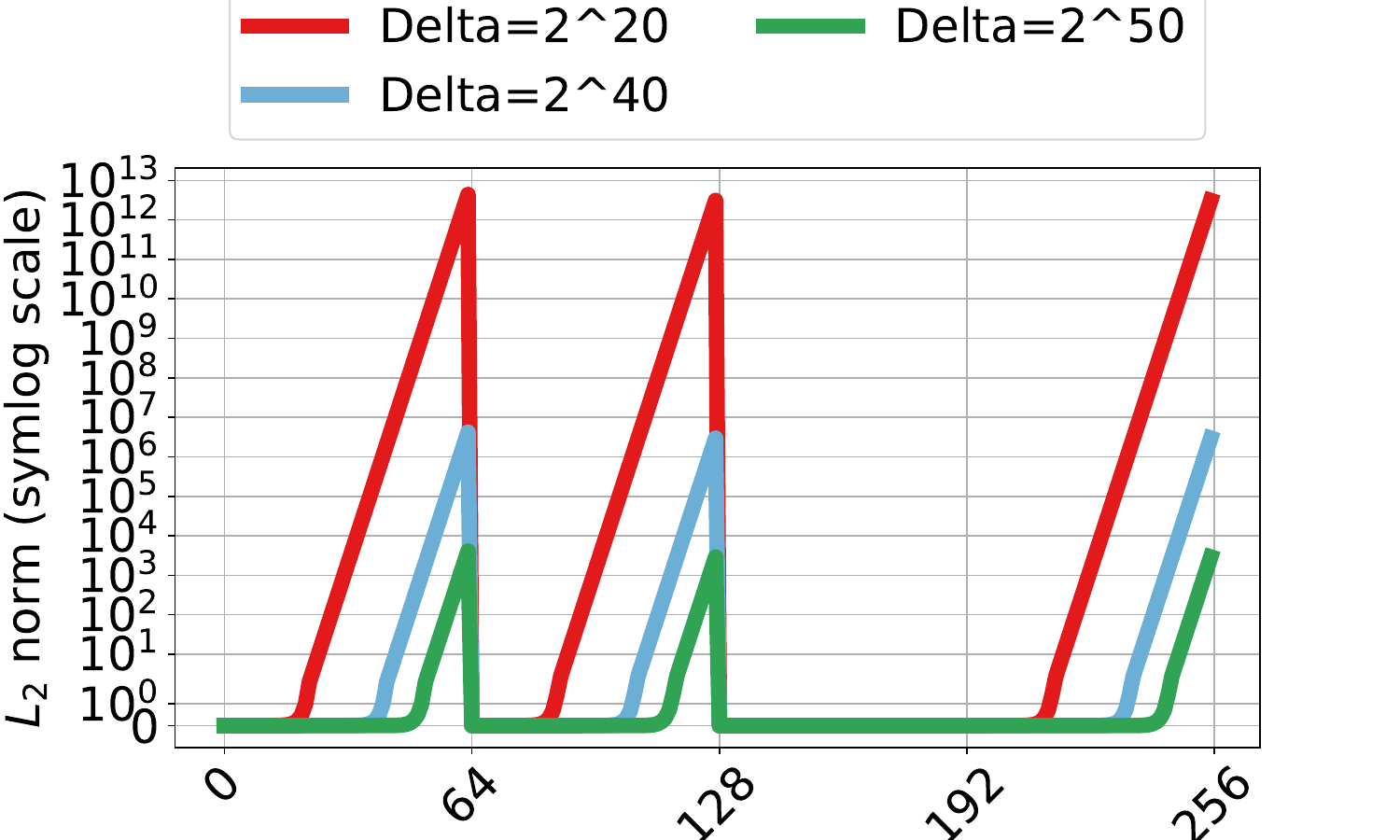}
    \caption{$L_2$-norm of the error after a single bit-flip in the plaintext during encoding, for various scaling factors $\Delta$.
    The horizontal axis indicates the bit position $j$; the vertical axis is on a semi-log scale.
    Ring dimension: $N=4$.}
    \label{fig:theoreticDeltas}
\end{figure}
As shown in Figure~\ref{fig:theoreticDeltas}, increasing $\Delta$ significantly reduces the error magnitude and increases the number of resilient bits.
This occurs because the impact of errors becomes noticeable only when the flipped bit corresponds to or exceeds the magnitude of $\Delta$.
In other words, the larger the scaling factor, the more initial bits remain unaffected, yielding smaller error norms and higher bit-level fault tolerance.

This empirical observation is confirmed analytically by equation~\ref{eq:norm2}, which shows that the error is inversely proportional to the scaling factor.
When an error $e_{i,j}$ is divided by $\Delta$, noise components with magnitudes smaller than the bit precision of $\Delta$ are effectively suppressed in the final result.
This mathematical relationship explains why larger scaling factors provide enhanced fault resilience: they act as a natural filter that attenuates the impact of small-magnitude bit flips on the decoded output.

However, this resilience comes at the cost of reduced fractional precision in the encoded message (see Section~\ref{section:background}).
The selection of $\Delta$ therefore requires careful consideration, balancing the competing demands of fault resilience and computational accuracy.

\subsection{Theoretical Estimation of Error during Encryption using Vanilla CKKS}
\label{subsec:plainvanilla}

Under public-key encryption, a vector $\mathbf{m}$ is encoded into a polynomial $\boldsymbol{m}(X)$, encrypted as in Equation~\ref{eq:cipherPublic}, and decrypted using Equation~\ref{eq:desencrypt}.

\subsubsection{Bit Flip in the Plaintext Before Encryption}\label{subsubsec:teoPlaintextPre}

Assuming that a bit flip introduces an error $\boldsymbol{e}_{i,j}$, as previously explained, the plaintext becomes:
\begin{equation}
\boldsymbol{m}' = \boldsymbol{m} + \boldsymbol{e}_{i,j}
\end{equation}

Encrypting $\boldsymbol{m}'$ using the public key (Equation~\ref{eq:cipherPublic}) yields:
\begin{equation}
  \begin{split}
    ct'
    &= \bigl([\,\boldsymbol{m}' + p_0v + e_1\,]_Q,\;[\,p_1v + e_2\,]_Q\bigr) \\
    &= \bigl([\,\boldsymbol{m} + \boldsymbol{e}_{i-1,j} + p_0v + e_1\,]_Q,\;c_1\bigr)
    = \bigl(c_0',\,c_1\bigr).
  \end{split}
\end{equation}

The error affects only the ciphertext component $c_0'$ in the coefficient $i$, with a maximum magnitude $|e|\le Q/2$ due to modular arithmetic.
Decryption (Equation~\ref{eq:desencrypt}) of such erroneous ciphertext yields:
\begin{equation}
\boldsymbol{m}'' = [\,c_0' + c_1\,s\,]_Q
     = \boldsymbol{m} + \boldsymbol{e}_{i-1,j} + (e_1 + e_2\,s)
\end{equation}

However, the error magnitude grows exponentially with the flipped bit position, as each bit represents a successive power of 2 in the binary representation.
Consequently, an exhaustive bit-flip analysis across all coefficient positions, with L2 norm computation for each perturbation, yields the exponential error scaling pattern illustrated in Figure~\ref{fig:theoreticNorm} (Section\ref{subsec:teoencdec}).
The bit-flip impact remains localized to the position $i$ and scales linearly with $e_j$.


\subsubsection{Bit Flip in the Ciphertext} \label{subsubsec:teoCipher}

If the bit flip occurs in the ciphertext, two sub-cases arise:
\paragraph{Flip in $c_0$}

For a perturbed ciphertext $ct'=(c_0+\boldsymbol{e}_{i-1,j},\;c_1)$, the behavior mirrors the previous plaintext case: the error remains localized in coefficient $i$ and results in a decoding error proportional to $e_j$.

\paragraph{Flip in $c_1$}

For a perturbed ciphertext $ct'=(c_0,\;c_1+\boldsymbol{e}_{i-1,j})$, decryption proceeds as:
\begin{equation*}
\begin{split}
    \boldsymbol{m}''&= [\,c_0 + (c_1 + \boldsymbol{e}_{i-1,j})\,s\,]_Q \\
    &= [\,(c_0 + c_1\,s) + (\boldsymbol{e}_{i-1,j}\,s)\,]_Q \\
     &= [\,(\boldsymbol{m}  + e_1 + e_2\,s) + (\boldsymbol{e}_{i-1,j}\, s)\,]_Q
\end{split}
\end{equation*}
where $\boldsymbol{e}_{i-1,j}\, s$ involves polynomial multiplication.
Thus, the error spreads across \textit{all} $N$ coefficients. Since the secret key $s$ has small coefficients (much less than $q_0$), each coefficient in the result incurs an error of magnitude proportional to $e_j$, but this error is now distributed.
The maximum magnitude is bounded by $|e_j|\cdot\|s\|_\infty$.

\subsubsection{Bit flip in the Plaintext Before Decoding}

If a bit flip is introduced in the coefficient $i$ of a plaintext $\boldsymbol{m}$, after decryption but before decoding, the effect is analogous to the case analyzed in Section~\ref{subsec:teoencdec}, since the plaintext being decoded is equivalent to:
\begin{equation}
\boldsymbol{m}' = \boldsymbol{m} +\boldsymbol{e}_{i-1,j}
\end{equation}
which is the same structure previously analyzed.

\subsection{Theoretical Estimation of Error in RNS} \label{subsec:teorns}
CKKS typically uses a Residue Number System (RNS) representation for the modulus  $Q = \prod_{k=0}^{L-1} q_k$, where $L$ determines the number of available levels (i.e., number of supported homomorphic operations).

Each coefficient $p\in\mathbb{Z}_Q$ is represented by its residues:
\begin{equation*}
r_k = p \bmod q_k \quad \text{for } k = 0, \dots, L-1
\end{equation*}

When applied to all coefficients of a polynomial (either in the plaintext or in the ciphertext), this yields $L$ \emph{residue polynomials}, commonly referred to as \emph{limbs}.
Reconstruction via the Chinese Remainder Theorem (CRT), as shown in Mohan~\cite[Chapter~5]{Mohan2016}, is:
\begin{equation}
p = \left(\sum_{k=1}^L r_k \left[\left(\frac{1}{Q_k}\right)\bmod q_k \right] Q_k \right) \text{ mod } Q   \label{eq:crt}
\end{equation}
where $Q_k = Q/q_k$ and $\frac{1}{Q_k}$ are the multiplicative inverse of $Q_k$.

A bit flip on the $j^{th}$ bit affecting the $k$-th residue changes it to $r_k' = r_k + e_{k,j}$, which introduces an error into the reconstructed value:
\begin{equation}
    p' - p= e_{k,j}\,Q_k\,\left[\left(\frac{1}{Q_k}\right)\bmod q_k \right]\pmod{Q}
\end{equation}

This error can be as large as $Q_k$, potentially spanning hundreds or thousands of bits, even though the bit flip originally affected a single bit in a single residue.

\subsection{Theoretical Estimation of Error in NTT}\label{subsec:teonttNeg}

To accelerate encoding and decoding operations, CKKS uses the negacyclic Number Theoretic Transform (NTT) over the ring $\mathbb{Z}_q[X]/(X^N+1)$.
Let $\psi\in\mathbb{Z}_q$ be a primitive $2N$-th root of unity satisfying:
\begin{equation*}
\psi^2\equiv \xi,\quad \psi^N\equiv -1\pmod{q},
\end{equation*}
where $\xi$ is a primitive $N$-th root of unity.
The NTT matrix is defined as:
\begin{equation*}
W_{NTT}=
\begin{psmallmatrix}
\psi^{2(0 \times 0) . 0} & \psi^{2(0 \times 1) . 1} & \cdots & \psi^{2(0 \times N-1) . N-1} \\
\psi^{2(1 \times 0) . 0} & \psi^{2(1 \times 1) . 1} & \cdots & \psi^{2(1 \times N-1) . N-1} \\
\psi^{2(2 \times 0) . 0} & \psi^{2(2 \times 1) . 1} & \cdots & \psi^{2(2 \times N-1) . N-1} \\
\vdots & \vdots & \ddots & \vdots \\
\psi^{2(N-1 \times 0) . 0} & \psi^{2(N-1 \times 1) . 1} &\cdots & \psi^{2(N-1 \times N-1) . N-1}
\end{psmallmatrix}
\end{equation*}

We denote its inverse as $W^{-1}_{NTT}$.
When a single bit flip occurs at bit position $j^{th}$ within coefficient $i$, the induced perturbation after NTT transformation becomes $\mathbf{m}' - \mathbf{m} =  W_{NTT}\,\mathbf{e}_{i,j}$.
Therefore, the $L_2$ norm of the error is:
\begin{equation*}
  \lVert  \mathbf{m}' - \mathbf{m}\rVert_2
  = \lVert W_{NTT}\,\mathbf{e}_{i,j}\rVert_2.
\end{equation*}

This behavior parallels that of the Discrete Fourier Transform (DFT), except that the NTT operates over $\mathbb{Z}_q$ with moduli $q$ typically of $30-60$ bits.
The twiddle factors, being roots of unity modulo $q$, can amplify errors during the transformation, thereby increasing the error.

\begin{figure}[!ht]
  \centering
  \includegraphics[width=0.8\linewidth]{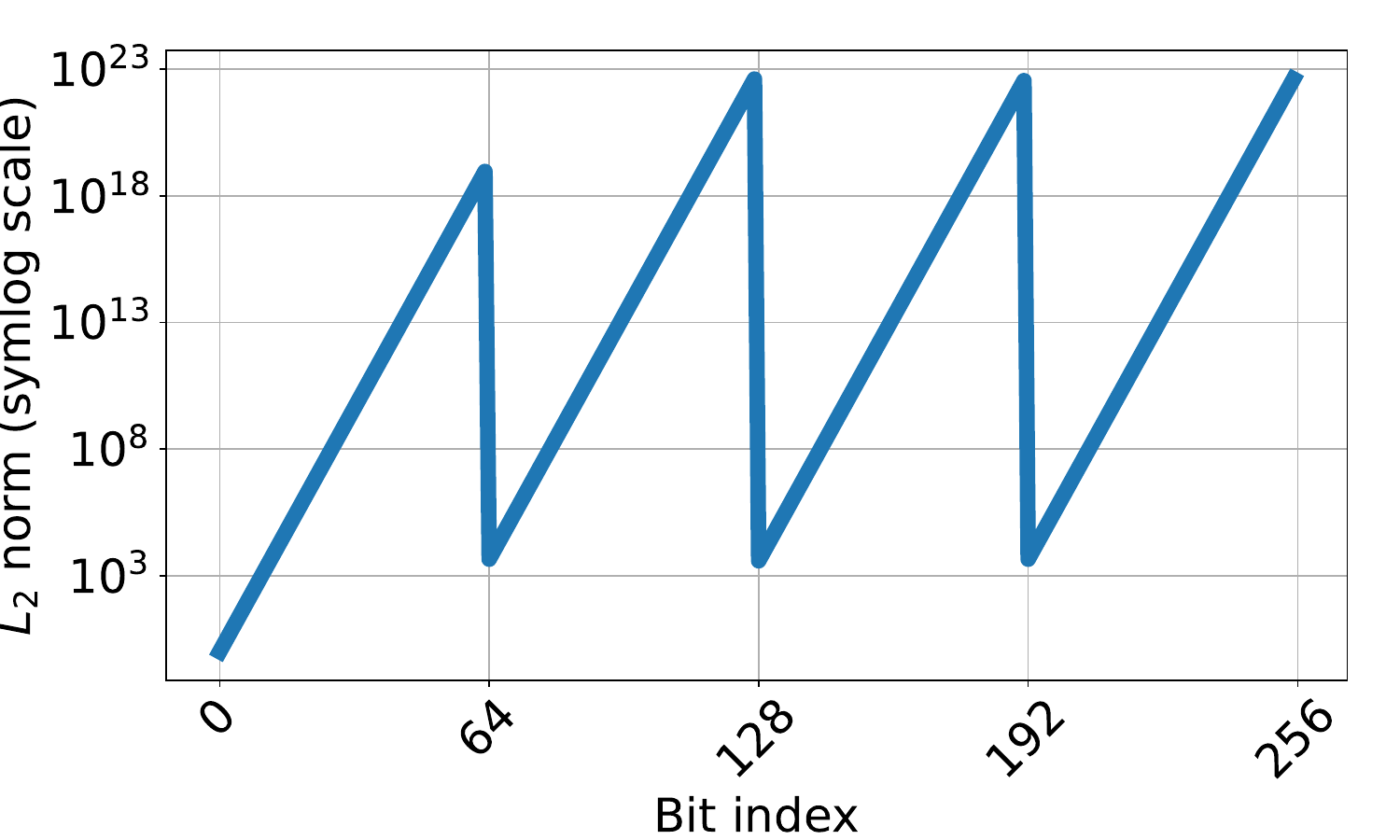}
  \caption{$L_2$-norm of the error after a single bit-flip in the plaintext during encoding.
  The horizontal axis shows the modified bit position $i \times 64 + j$; the vertical axis shows the $L_2(i,j)$ norm on a logarithmic scale. Ring dimension: $N=4$.}
  \label{fig:theoreticNormNTT}
\end{figure}

Figure~\ref{fig:theoreticNormNTT} shows the values of $L_2(i,j)$ for $N=4$ and 64-bit coefficients in the NTT case.
The horizontal axis indicates the bit-flip position $i \times 64 + j$, and the vertical axis shows $L_2(i,j)$ on a logarithmic scale.
As Figure~\ref{fig:theoreticNormNTT} shows, even flips in the least significant bits cause the norm (error) to exceed four orders of magnitude--producing errors large enough to make correct decoding practically impossible, as will be further illustrated in Section~\ref{section:results}.

\subsection{Summary}\label{subsec:summary}
\paragraph*{Bit flip in the plaintext before encryption} the error remains localized in a single coefficient (the one where the bit flip occurred), with a magnitude proportional to $e_j=2^j$, depending on the position $j$ of the flipped bit.
\paragraph*{Bit flip in the ciphertext ($c_0$)} identical to the plaintext case--the error remains localized after decryption, with magnitude proportional to the bit position $j$.
\paragraph*{Bit flip in the ciphertext ($c_1$)} the error is multiplied by the secret key $s$, becoming distributed across \textit{all} $N$ coefficients; its maximum magnitude is $\mathcal{O}(|e_j|\cdot\|s\|_\infty)$.
\paragraph*{Bit flip in the plaintext before decoding} behaves like a localized error in a single coefficient of the encoded polynomial--effectively equivalent to the encoding case, plus minor encryption noise.
\paragraph*{RNS} a bit flip in a single \emph{limb} (residue) can result in a reconstruction error of order $\mathcal{O}(Q_k)$, where $Q_k=Q/q_k$ is typically a large value.
\paragraph*{Negacyclic NTT} a bit flip before or after the NTT produces an error proportional to $W_{NTT}\,\mathbf{e}_{i,j}$, spreading across all coefficients. Entries of $W_{NTT}$ are bound by modulus $q$, which typically ranges from \SIrange{30}{60}{bits}, thus potentially amplifying the error significantly.


\section{Methodology}\label{section:methodology}

This section describes the methodology used to experimentally quantify the resilience of the CKKS scheme under single bit-flip fault injections.

\subsection{Scheme Definition and Operational Assumptions}

We consider four experimental configurations using CKKS:

\begin{itemize}
  \item \textbf{Vanilla mode:} All low-level performance optimizations are disabled; arithmetic occurs in a single large modulus in coefficient space (no RNS--single limb--and no NTT domain representation).
  \item \textbf{RNS-only mode:} RNS decomposition is enabled (multi-limb representation), while NTT is bypassed.
  \item \textbf{NTT-only mode:} NTT is enabled to accelerate polynomial multiplication / convolution, while RNS decomposition is disabled (single-limb modulus).
  \item \textbf{RNS+NTT mode:} Fully optimized configuration with both RNS decomposition and NTT transform.
\end{itemize}

This setup allows us to isolate the individual and combined effects of RNS and NTT optimizations on fault propagation and resilience, while maintaining compatibility with standard security parameters.

In preliminary tests, we observed that both the \textit{NTT-only} and the \textit{RNS+NTT} configurations exhibit similar error amplification even for single-bit faults, in line with our theoretical analysis (Section~\ref{section:error-resilience-analysis}).  In other words, a single bit flip at any stage of these two cases leads to undecryptable results.
As a result, we focus our analysis on the other two cases (\textit{vanilla} and \textit{RNS-only} modes), where more interesting behavior can be observed and meaningfully characterized.

\subsection{Fault / Error Injection Model}

As in our theoretical analysis (Section~\ref{subsec:teosingleBitFlip}), we assume a transient single-bit fault model, motivated by soft errors such as radiation-induced bit flips in processor registers or memory buffers.
Specifically, we inject a single bit flip in one coefficient of the output of a given operation during each execution.
In the case of \texttt{Encode} and \texttt{Decode}, we inject the fault into the plaintext representation; for \texttt{Encrypt} and \texttt{Decrypt}, we inject it into the ciphertext.
This approach models a fault affecting the destination register or output memory of these functions, without modifying the inputs or intermediate computations.
\begin{figure}[!ht]
    \centering
    \includegraphics[width=0.9\linewidth]{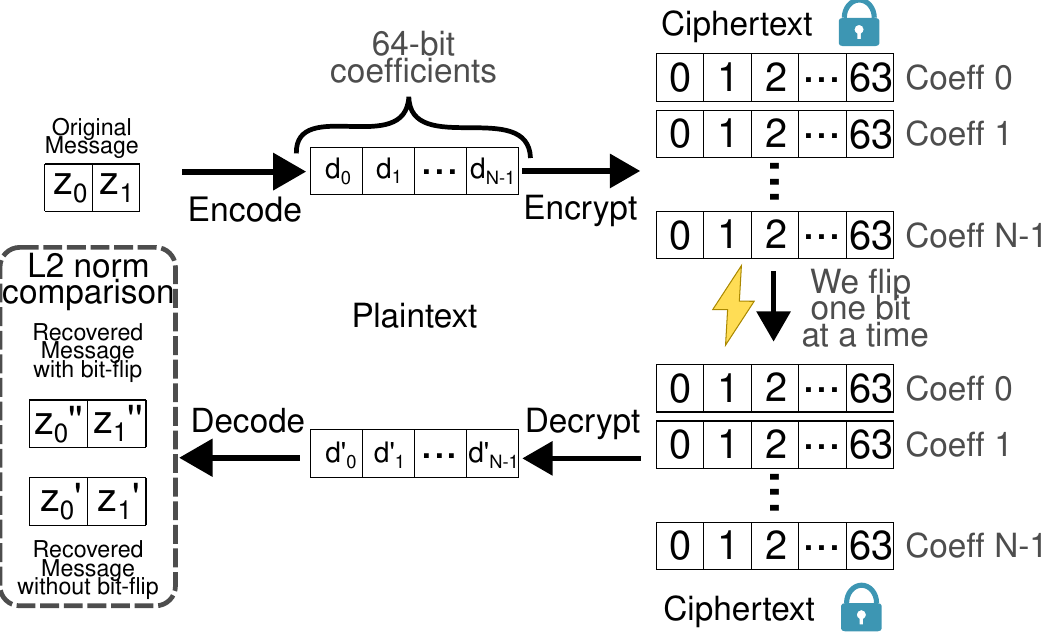}
    \caption{An input is first encoded and then encrypted. A single bit flip is applied to one coefficient at a time and at one bit position. The ciphertext is then decrypted and decoded, and the resulting output is compared against the unmodified reference output using the \(L_2\)-norm. For illustration, the ciphertext is represented as a single polynomial.
    }
    \label{fig:metodologyInjection}
\end{figure}
Figure~\ref{fig:metodologyInjection} illustrates how the fault is injected at a single stage per experiment, directly on the polynomial representation (plaintext or ciphertext), simulating a post-operation corruption that propagates through subsequent stages of the computation.

\subsection{Implementations and Instrumentation}

\subsubsection{CKKS Libraries}
We use two different CKKS implementations: \texttt{HEAAN 1.0}.
First publicly available CKKS implementation. This specific version does not support RNS and NTT, which makes it suitable for the evaluation of the \textit{vanilla mode}. It uses arbitrary-precision coefficients managed by the \texttt{NTL} library.
\texttt{OpenFHE 1.3.0}: highly-optimized open-source (de facto) standard implementation of CKKS~\cite{AlBadawi2022}. It natively supports RNS and NTT, using 64-bit internal coefficients.

We minimally adapt both libraries to gain complete control over the selection of pseudo-random number generator (PRNG) seeds. This ensures strict reproducibility of experiments and allows precise isolation of the effects caused by individual bit flips.

In the case of \texttt{HEAAN~1.0}, we consider the following implementation details:
\begin{itemize}
    \item The bit length of each plaintext coefficient is not fixed in advance; it depends on internal combinations of $q_0$ and $\Delta$, and may vary depending on the PRNG seed used by the scheme.
    \item The ciphertext bit length is of length $q_0$.
    \item Due to NTL implementation constraints, bit positions larger than the $Q$ of the coefficient are ignored in internal operations, which limits the scope of bit-level analysis.
\end{itemize}

In the case of \texttt{OpenFHE}, we consider the following implementation details:
\begin{itemize}
    \item Use of 64-bit integer coefficients (\texttt{uint64\_t}) without arbitrary-precision arithmetic.
    \item The client can configure the initial modulus $q_0$.
    This parameter determines the starting bit-level precision of the scheme. The subsequent moduli $q_i$ are automatically selected to match the bit length of $q_0$, ensuring uniform precision across levels.
    \item  The plaintext and the ciphertext belong to the same ring.
    They both have polynomial coefficients of size $q_0$, regardless of the scaling factor $\Delta$.
    \item The total number of moduli depends on the required multiplicative depth of the computation. One moduli for each RNS limb.
    \item 128-bit arithmetic support, which--though not used in this work--offers a pathway for future experiments.
\end{itemize}

\subsubsection{Using OpenFHE as Vanilla CKKS Implementation}
\label{subsec:emulacion}

\texttt{OpenFHE} is substantially more computationally efficient than \texttt{HEAAN}.
Consequently, we primarily use \texttt{OpenFHE} in this work, although selected experiments are replicated in \texttt{HEAAN} for validation purposes.
However, it does not natively offer a mechanism to disable RNS and NTT optimizations to enable \texttt{OpenFHE} on vanilla mode.
We explicitly bypass both RNS and NTT optimizations.
\textbf{RNS}: We enforced a single-limb representation, which is feasible for depth-zero operations (i.e.,~no ciphertext multiplications).
Additionally, automatic rescaling was disabled and replaced with manual scaling to prevent the introduction of additional limbs.
\textbf{NTT}: Before injecting a bit flip, we move the polynomial from evaluation space back to coefficient space using the inverse NTT (INTT). We flip the desired bit and transform the polynomial back to the evaluation space using NTT. After this hook, the pipeline continue normally.

Since the NTT operates modulo $q_0$, flipping bits beyond the bit width of $q_0$ causes modular wraparound, which can alter the intended effect of the bit flip and potentially lead to catastrophic decryption errors. To minimize distortion in the analysis, we set $q_0$ to the maximum value supported by the library--60 bits--ensuring that most bit flips occur within the valid range and reducing unintended side effects from modular arithmetic.

\subsubsection{Validation of Vanilla Mode using \texttt{OpenFHE}}

To ensure that vanilla \texttt{OpenFHE} mode faithfully replicates the behavior of \texttt{HEAAN~1.0}, we follow these validation steps, 1) Configure \texttt{OpenFHE} on vanilla mode (disabling RNS and NTT) and set $q_0$ and $\Delta$ to match \texttt{HEAAN}'s effective bit precision, 2) Restrict input values in \texttt{HEAAN} to real numbers, in line with \texttt{OpenFHE}'s input constraints, 3) Compare $L_2$-norms and error statistics after identical bit flips are introduced in both libraries and 4) Verify consistent sensitivity patterns across low- and high-order bit positions.

\subsection{Experimental Procedure}

In \texttt{OpenFHE}, ciphertexts and plaintexts are internally represented using NTT transformation (evaluation domain). Since we need to inject faults in the coefficients of the polynomials, we must ensure that we are working in the coefficient domain. For the vanilla and RNS modes, we first apply the inverse NTT to bring the polynomial back to the coefficient domain, then we perform the bit flip, and finally we reapply the NTT to restore the polynomial to the evaluation domain.

For each bit flip experiment, we follow these steps:
\begin{enumerate}
    \item Encode and, if applicable, encrypt the input vector using the selected library.
    \item Apply the inverse NTT (INTT) to bring the polynomial into coefficient space.
    \item Flip a single bit within one coefficient (64-bit precision in \texttt{OpenFHE}; arbitrary-precision in \texttt{HEAAN}).
    \item Reapply NTT to return the polynomial to evaluation space.
    \item Proceed with the homomorphic operation pipeline.
    \item Decrypt and decode the result.
    \item Compute evaluation metrics.
\end{enumerate}

We repeat this process for every bit position in each coefficient of the different stages of the CKKS pipeline.

\paragraph{Input Vector Generation and PRNG Control} \label{subsec:prng}

Our analysis is application-agnostic, employing input vectors of random real numbers drawn from a uniform distribution.
To assess the sensitivity of the CKKS scheme, we sample inputs from various intervals whose boundaries are defined by powers of two. Unless otherwise specified, input values are sampled as uniformly distributed real numbers within the interval \numrange{0}{256}, and represented in double-precision floating-point format. The bounded input range enables testing multiple scaling factors ($\Delta$), allowing for a more comprehensive analysis of their impact on encoding and rounding errors.
We analyze the trade-off between precision and noise growth by varying $\Delta$ across a wide range, ensuring it stays within the coefficient bounds detailed in Section~\ref{sec:ckks-parameters}.

To evaluate stability impact of the seed variation, we repeat each experiment (i.e.,~each bit flip) 2500 times.
Each repetition involves 100 different seeds for the scheme's PRNG and 25 for the input generator.
These seed combinations are fixed and shared across all experiments, ensuring a fair comparison using equivalent input conditions.
The variability observed across repetitions is negligible, with no discernible differences in the resulting patterns, indicating that the adopted sampling strategy is sufficient for the purposes of this study.

\paragraph{Number of Coefficients}
In order to ensure reasonable computation times given the number of seeds, we first studied the effect of the ring dimension $N$ on bit-flip sensitivity patterns.
This parameter has the greatest impact on computational performance: a larger $N$ provides better security, but at the cost of increased computation.
Since the bit-flip sensitivity patterns were found to be consistent across experiments with varying ring dimensions $N$ (from $2^3$ up to $2^{16}$), we restrict our analysis to smaller values of $N$.
This choice both improves performance and facilitates visualization: it is far easier to discern structure in the behavior of a small set of coefficients in a single plot than to interpret patterns spread across thousands simultaneously.

\paragraph{Analyzed Stages}
As presented in Section~\ref{section:error-resilience-analysis}, we corroborate experimentally that flipping a single bit in the plaintext during the \texttt{Encode} or \texttt{Decode} stages has the same effect as flipping a bit in the first polynomial ($c_0$) of the ciphertext during \texttt{Encrypt} or \texttt{Decrypt}.
For this reason, we focus our study on the behavior of the ciphertext under bit-flip faults.
This approach allows us, within a single experiment, to analyze both the behavior of the first polynomial (and thus the plaintext) and the behavior of the second polynomial of the ciphertext.

\subsection{Evaluation Metrics} \label{subsec:metricas}

Given the original decrypted vector $\mathbf{x}$ (without bit flips) and the resulting vector $\mathbf{y}$ after error injection and decryption, we quantify the deviation introduced by each bit flip by two complementary evaluation metrics:

\paragraph{Mean Squared Error (MSE)}
We define the element-wise error as $\delta_i = y_i - x_i$, and the MSE is computed as:
\begin{equation}\label{eq:MSE}
\mathrm{MSE} = \sqrt{\frac{1}{k}\sum_{i=1}^{k} \delta_i^2}
\end{equation}

\paragraph{Element-wise Relative Error}
For each decrypted element $i$, we compute the relative error:
\begin{equation}\label{eq:relativeError}
\epsilon_i = \frac{|y_i - x_i|}{|x_i|} =  \frac{|\delta_i|}{|x_i|}
\end{equation}

To determine whether a vector is correctly recovered, we use the metric in Eq.~\eqref{eq:relativeError} and proceed element-wise.
For each element \(i\), we compute the relative error with the reference value (i.e,~same element of the vector recovered wihout no bit flip).
We then classify the outcome of the procedure according to the fraction of elements correctly recovered:

\begin{itemize}
  \item \textbf{Catastrophic}: less than \SI{1}{\percent} of vector elements are correctly recovered
  \item \textbf{Application-dependent}: \SIrange{1}{99}{\percent} of vector elements are correctly recovered
  \item \textbf{Robust}: more than \SI{99}{\percent} of vector elements are correctly recovered
\end{itemize}

\subsection{Experimental Setup}
\label{subsec:setup}

All experiments are execute on a dedicated workstation equipped with an Intel Core i7-11700 CPU (8 physical cores, 16 threads, 16\,MB L3 cache, base 2.50\,GHz) and 32\,GB DDR4 RAM (dual channel, 3200\,MHz).
The operating system is Arch Linux (kernel \texttt{Linux 6.9.7-arch1-1}, distribution snapshot \texttt{257.5-1}).
We compiled both \texttt{OpenFHE 1.3.0} and \texttt{HEAAN 1.0} in release mode using \texttt{g++ 15.2.1} with flags \texttt{-O3}.
For \texttt{HEAAN}, NTL (\texttt{NTL 11.5.1}) was linked dynamically; \texttt{OpenFHE} was built with its default backend (no GPU acceleration).

\section{Experimental Results}\label{section:results}

This section presents the results obtained from injecting single bit flips during the encryption stage of the CKKS scheme, as described in Section~\ref{section:methodology}, in order to evaluate its resilience against induced faults.

\subsection{Vanilla CKKS}\label{ssec:ckks-vanilla}
We first analyze the \emph{vanilla} CKKS scheme using \texttt{OpenFHE}; i.e., without RNS and NTT optimizations.

\subsubsection{Evolution of the $L_2$-norm Under Single Bit Flips}\label{subsec:encN2}
To evaluate the fault sensitivity of the \textit{vanilla} CKKS variant, we systematically flipped one bit at a time in the ciphertexts generated during encryption, covering all bit positions in both $c_0$ and $c_1$ polynomials.
After each injection, the modified ciphertext was decrypted and decoded, and the $L_2$-norm of the difference with respect to the fault-free output was computed.
This metric quantifies the magnitude of the decoding error associated with each bit position.

\begin{figure}[!ht]
    \centering
    \includegraphics[width=0.8\linewidth]{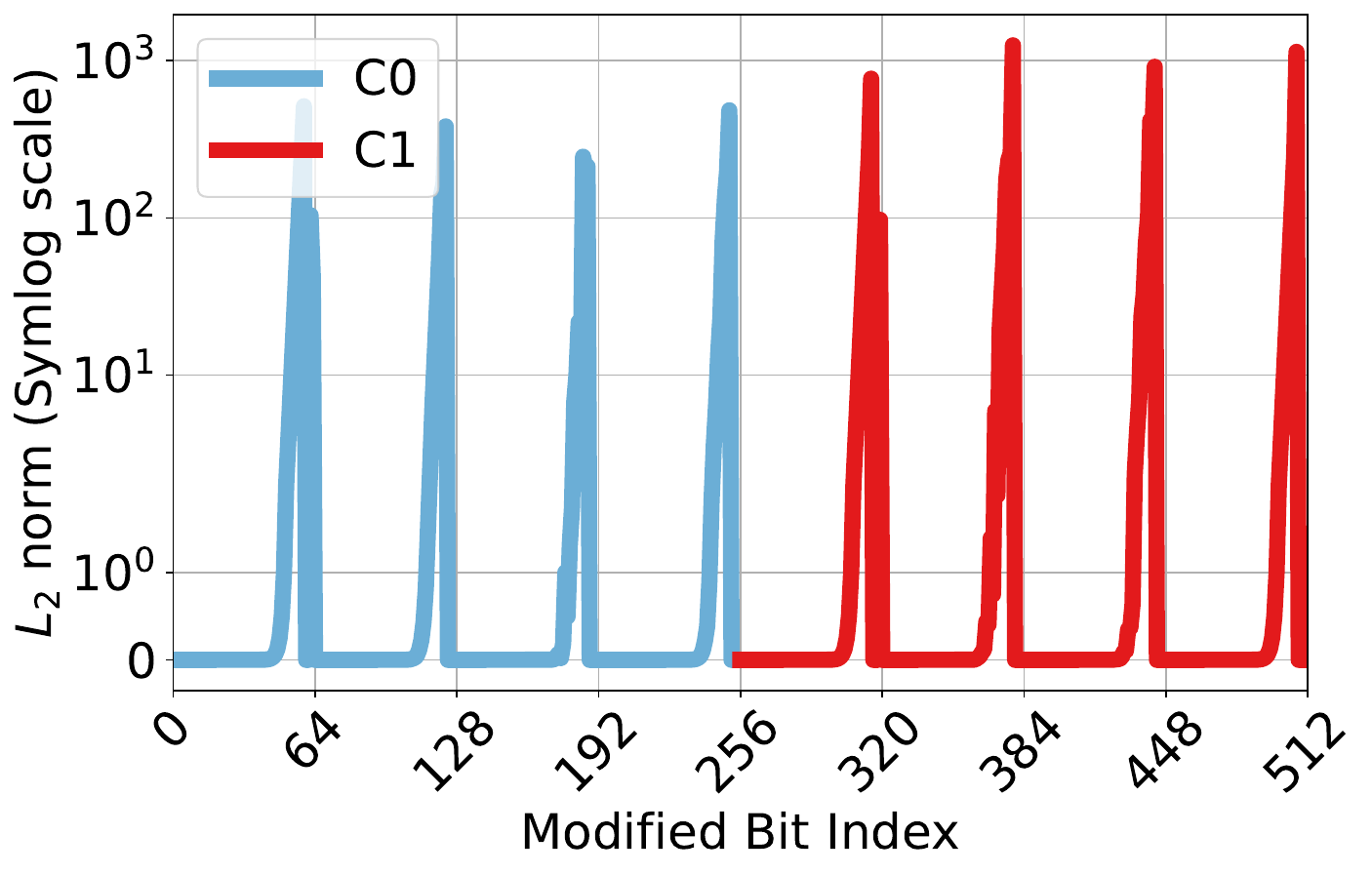}
    \caption{
        Bit flip during encryption on vanilla mode. The horizontal axis indicates the position of the flipped bit within the ciphertext, while the vertical axis shows the resulting $L_2$-norm of the decoding error. Parameters: 60‑bit $q_0$, input range from \numrange{0}{256},  $\Delta = 2^{50}$, and  $N = 4$.}
    \label{fig:EncryptNormOpenFHE}
\end{figure}

Figure~\ref{fig:EncryptNormOpenFHE} shows the resulting $L_2$-norm distribution for all possible single-bit flips, highlighting which regions of the ciphertext are most sensitive to faults.
The ring dimension is $N=4$, with modulus $q_0=60$-bit and scaling factor of $\Delta=2^{50}$.
We observe that, for the first polynomial $c_0$, the behavior closely follows the theoretical trend shown in Figure~\ref{fig:theoreticNorm}: least significant bits exhibit higher resilience to bit flips, while the error magnitude grows exponentially from the bit position corresponding to $\Delta$ up to the bit position defined by the modulus $q_0$, after which it abruptly drops to zero.

In our theoretical analysis, as well as in \texttt{HEAAN}, the coefficient $N/2$ of $c_0$ is completely robust.
However, in \texttt{OpenFHE}, as we can observe in Figure~\ref{fig:EncryptNormOpenFHE} near bit 192, this behavior does not occur.
This difference arises because \texttt{OpenFHE}, during decoding, injects Gaussian noise into the plaintext prior to the DFT (a specialized FFT), perturbing both real and imaginary components as a countermeasure against key-recovery attacks.

The injected noise has a standard deviation proportional to the spread of the imaginary part of the pre-DFT vector (plaintext), which, due to the scheme’s properties, follows a Gaussian distribution.
Flipping any bit in any coefficient, including $N/2$, alters this spread; higher-order bit flips have a stronger effect, increasing the overall noise variance and propagating it across all polynomial coefficients, thereby removing the theoretical robustness of coefficient $N/2$. Disabling this noise injection in \texttt{OpenFHE} restores the expected behavior: coefficient $N/2$ of $c_0$ becomes robust to bit-flips.
Since \texttt{HEAAN~1.0} does not include this security noise, its experimental sensitivity matches the theoretical prediction exactly.

\subsubsection{Impact of the Scaling Factor} \label{subsec:encScale}
One of the key parameters of the CKKS scheme is the scaling factor~$\Delta$.
As analyzed in Section~\ref{section:error-resilience-analysis}, increasing~$\Delta$ reduces the relative impact of a single‐bit flip during encoding or encryption and also increases the proportion of bits that remain resilient.

\begin{figure}[!ht]
  \centering
  \includegraphics[width=0.8\linewidth]{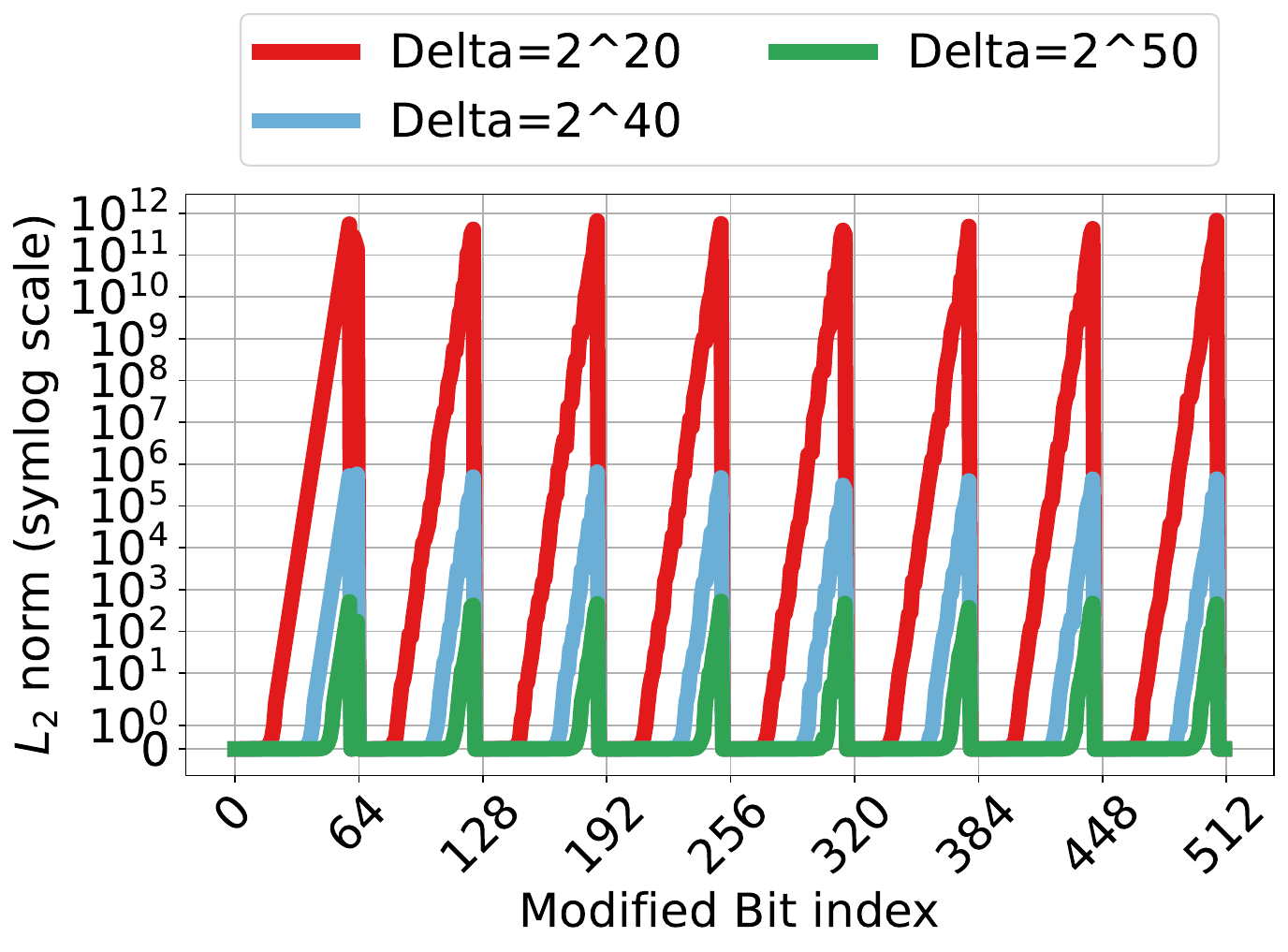}
  \caption{Bit flip during encryption on vanilla mode. The horizontal axis indicates the bit position in ciphertext \(c_0\), while the vertical axis shows the resulting $L_2$-norm of the decoding error on a symmetric log scale. Parameters: 60‑bit $q_0$, $\Delta \in \{2^{20},  2^{40}, 2^{50}\}$, input range from \numrange{0}{256} and  \(N=8\).}
  \label{fig:EncryptDeltasOpenFHE}
\end{figure}

To validate this behavior experimentally, we analyze the impact of three different scaling factors, namely $\Delta \in \{2^{20},  2^{40}, 2^{50}\}$.
Figure~\ref{fig:EncryptDeltasOpenFHE} shows the $L_2$-norm of the decoding error for a bit flip in the first ciphertext polynomial \(c_0\) on vanilla mode.
For clarity, only results for \(c_0\) are shown, since the behavior in \(c_1\) is analogous (as is shown in Fig.~\ref{fig:EncryptNormOpenFHE}).
We observe that higher values of~$\Delta$ yield lower $L_2$-norms after a bit flip, indicating stronger attenuation of induced errors. Furthermore, the point at which the error norm begins its exponential growth shifts in alignment with the magnitude of~$\Delta$, confirming that larger scaling factors increase the number of error resilient bits in each coefficient.
\begin{figure}[!ht]
  \centering
  \includegraphics[width=0.8\linewidth]{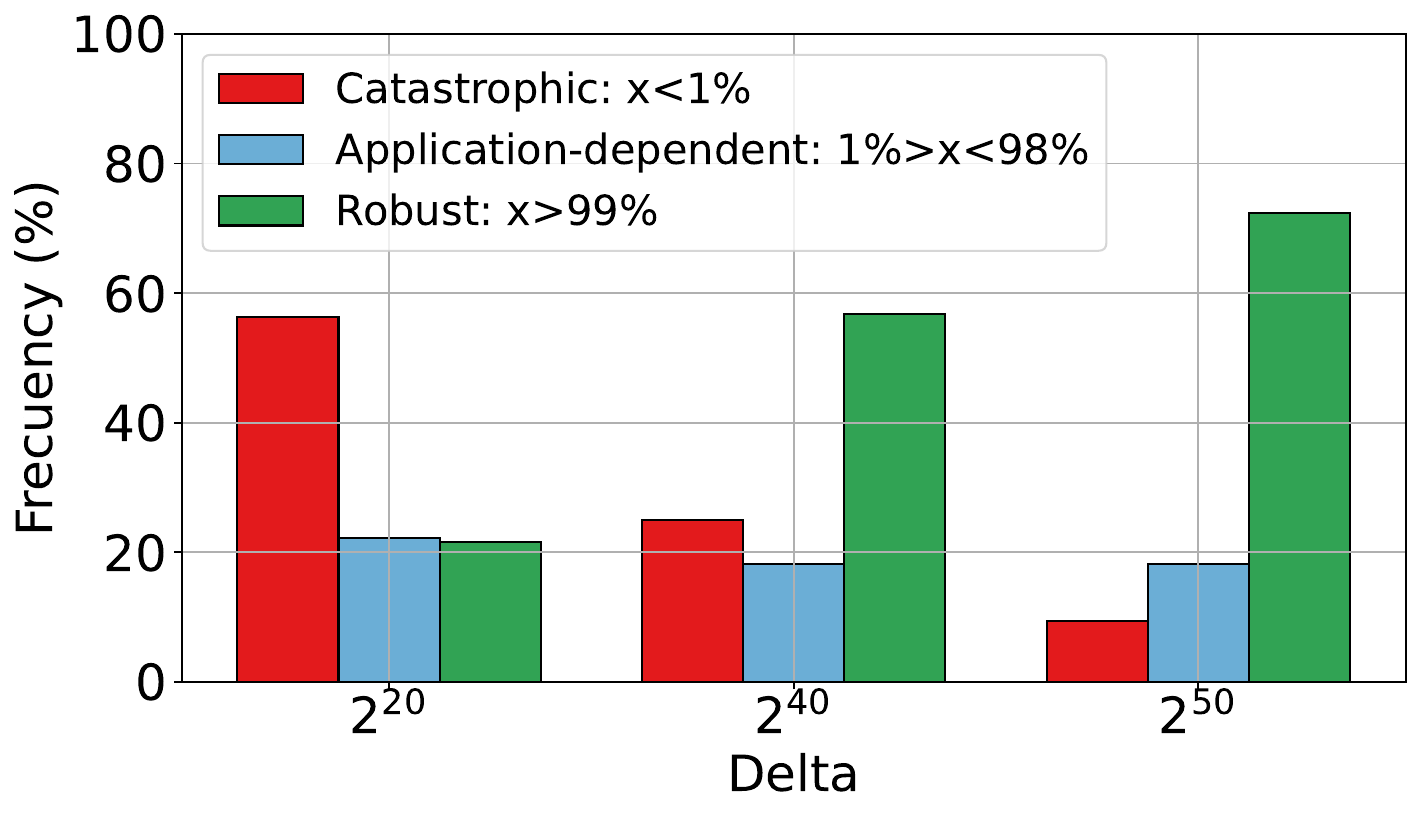}
  \caption{Histogram of error‐categorization outcomes for different scaling factors. Each group of bars corresponds to a specific \(\Delta\), showing the percentage of trials in the different categories.
  Parameters: 60‑bit $q_0$,  $\Delta \in \{2^{20}, 2^{40}, 2^{50}\}$, input range from \numrange{0}{256} and \(N= 2^{13}\).}
  \label{fig:EncryptCategoryMultHistOpenFHE}
\end{figure}
In Fig.~\ref{fig:EncryptCategoryMultHistOpenFHE}, we adopt a more realistic configuration using a ring dimension \(N = 2^{13} = 8192\) (4096 encryptable elements).
For each scaling factor \(\Delta \in \{2^{20}, 2^{40}, 2^{50}\}\) and each single-bit flip, we track the number of elements that decrypt correctly.
We use the three categories defined in Sec.~\ref{subsec:metricas} based on element-wise relative error (Eq.~\eqref{eq:relativeError}).
As \(\Delta\) increases, the frequency of \textbf{Robust} outcomes (green bars) also increases, confirming that a larger scaling factor provides greater tolerance to faults.


Although we do not explicitly compute the $L_2$-norm threshold at which decryption completely fails, the observed consistency of the \textbf{Application-dependent} error category across scaling factors indicates that the width of the transition interval—from full resilience to total failure—remains approximately constant for all values of \(\Delta\).
Within this transition region, some elements decrypt correctly while others fail, reflecting the gradual loss of robustness rather than an abrupt threshold.

We observe a slight increase in the \textbf{Application-dependent} category for $\Delta = 2^{20}$ and more markedly for smaller $\Delta$ (data not shown).
This is consistent with the chosen parameters: with $q_0=\SI{60}{\bit}$ and $\Delta=2^{20}$ the scheme provides roughly 20 fractional bits and about 40 integer bits.
When input values are limited to the range up to $2^{8}=256$, most of the available integer range remains unused, so small perturbations introduced during encoding/decoding (rounding, or bit-flip noise) can move some elements to the \textbf{Application-dependent} category.

\subsubsection{Impact of the Input Value's Magnitude} \label{subsec:encRange}

We fix $\Delta = 2^{20}$ and $q_0 = \SI{60}{\bit}$ to enable a wider representable input range.
As previously explained, the number of bits available to represent the integer part of an input value is approximately $q_0 - \log_2(\Delta)$--about 40 bits in this configuration.
In our experiments, we define input value intervals, from which we uniformly sample the elements.

\begin{figure}[!ht]
  \centering
    \includegraphics[width=0.8\linewidth]{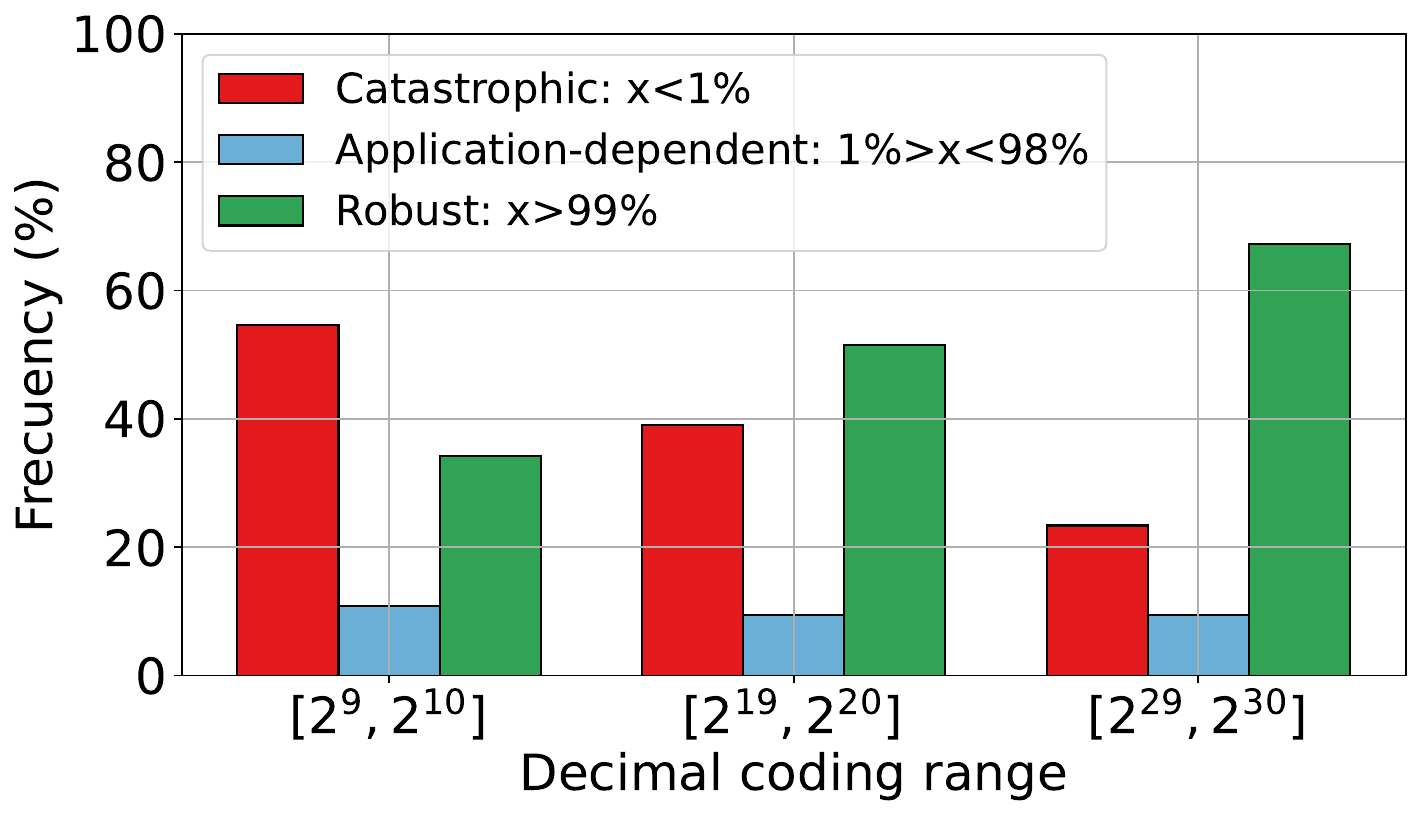}
    \caption{Histogram of error‐categorization outcomes for different scaling inputs. Each group of bars corresponds to a specific input range, showing the percentage of trials in the different categories.
  Parameters: 60‑bit $q_0$,  $\Delta=2^{20}$, input range $\in [2^{9},2^{10}], [2^{19},2^{20}]$ and $[2^{29},2^{30}]$ and  \(N= 2^{13}\).}
    \label{fig:EncryptCategoryHistOpenFHE_input}
\end{figure}

Figure~\ref{fig:EncryptCategoryHistOpenFHE_input} shows histograms of the error category distributions for three input intervals: $[2^{9},2^{10}]$, $[2^{19},2^{20}]$, and $[2^{29},2^{30}]$.
We deliberately avoid the interval $[2^{39},2^{40}]$ because it may overflow and cause a failed decryption even with unmodified inputs (i.e.,~without bit flips).
The selected intervals are representative of the range of input magnitudes considered in our experiments.
We use the following parameters: ring dimension $N = 8192$, ciphertext modulus $q_0 = \SI{60}{\bit}$, and scaling factor $\Delta = 2^{20}$.
Figure~\ref{fig:EncryptCategoryHistOpenFHE_input} presents a bimodal \textit{all-or-nothing} pattern: more than \SI{90}{\percent} of the cases, falls in robust or catastrophic categories.
This behavior is perceived also in Figure~\ref{fig:EncryptCategoryMultHistOpenFHE} but clearly show in this experiment.
Increasing the magnitude of the input values expands the set of elements in the robust category while reducing the fraction of catastrophic decryptions.
This is simply because the least significant bits of each coefficient only affect the least significant bits of the final result.
Therefore, with larger input values, their impact becomes relatively smaller.

\subsubsection{Effect of Reducing the Number of Encrypted Elements}\label{subsec:encGap}

\begin{figure}[!ht]
    \centering
    \includegraphics[width=0.8\linewidth]{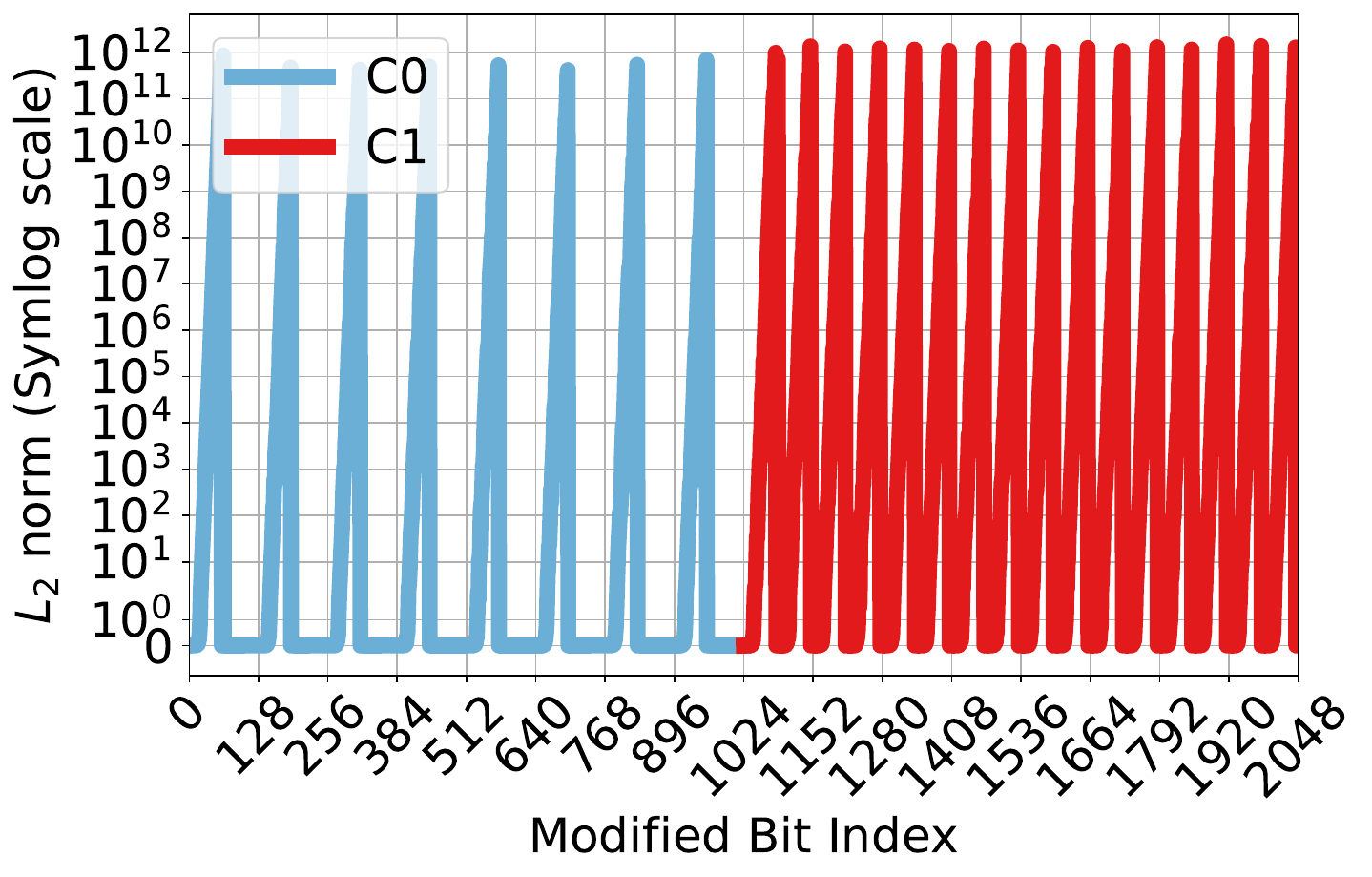}
    \caption{
        Bit flip during encryption on vanilla mode. The horizontal axis indicates the bit position in the ciphertext, while the vertical axis shows the resulting $L_2$-norm of the decoding error. Parameters: 60‑bit $q_0$,  \(\Delta=2^{20}\), sample input range from \numrange{0}{256},  \(N=16\), and four encrypted elements.}
    \label{fig:EncryptNorm_openfhe_gap}
\end{figure}

As noted previously, CKKS can encode up to \(N/2\) elements, where \(N\) is the power-of-two ring degree. When we encode fewer elements--still a power of two--we observe that additional coefficients remain unaffected by single‐bit flips.
Figure~\ref{fig:EncryptNorm_openfhe_gap} shows results for \(N=16\) with only \(N/4 = 4\) encrypted elements.
The results reveal that odd‐indexed coefficients of polynomial \(c_0\) remain unchanged under bit flips, whereas even‐indexed coefficients exhibit the same sensitivity patterns observed earlier.

As explained in Section~\ref{section:background}, the ratio between the maximum number of encryptable elements (\(N/2\)) and the number actually used defines an interval (or \textit{gap}) between the coefficients participating in the inverse transform. During the special FFT, only coefficients which indices are multiple of this gap are considered and scaled in the input vector (see Section~\ref{section:introduction} for details).
For example, when encoding \(N/2\) elements, all \(N\) coefficients of \(c_0\) are sensitive to bit flips, but encoding \(N/4\) elements reduces the number of sensitive coefficients of \(c_0\) to \(N/2\).
In the case of \(c_1\), all coefficients are sensitive to bit changes.

\subsection{Resilience in RNS-only mode}\label{subsec:encRNS}
\begin{figure}[!ht]
    \centering
    \includegraphics[width=0.8\linewidth]{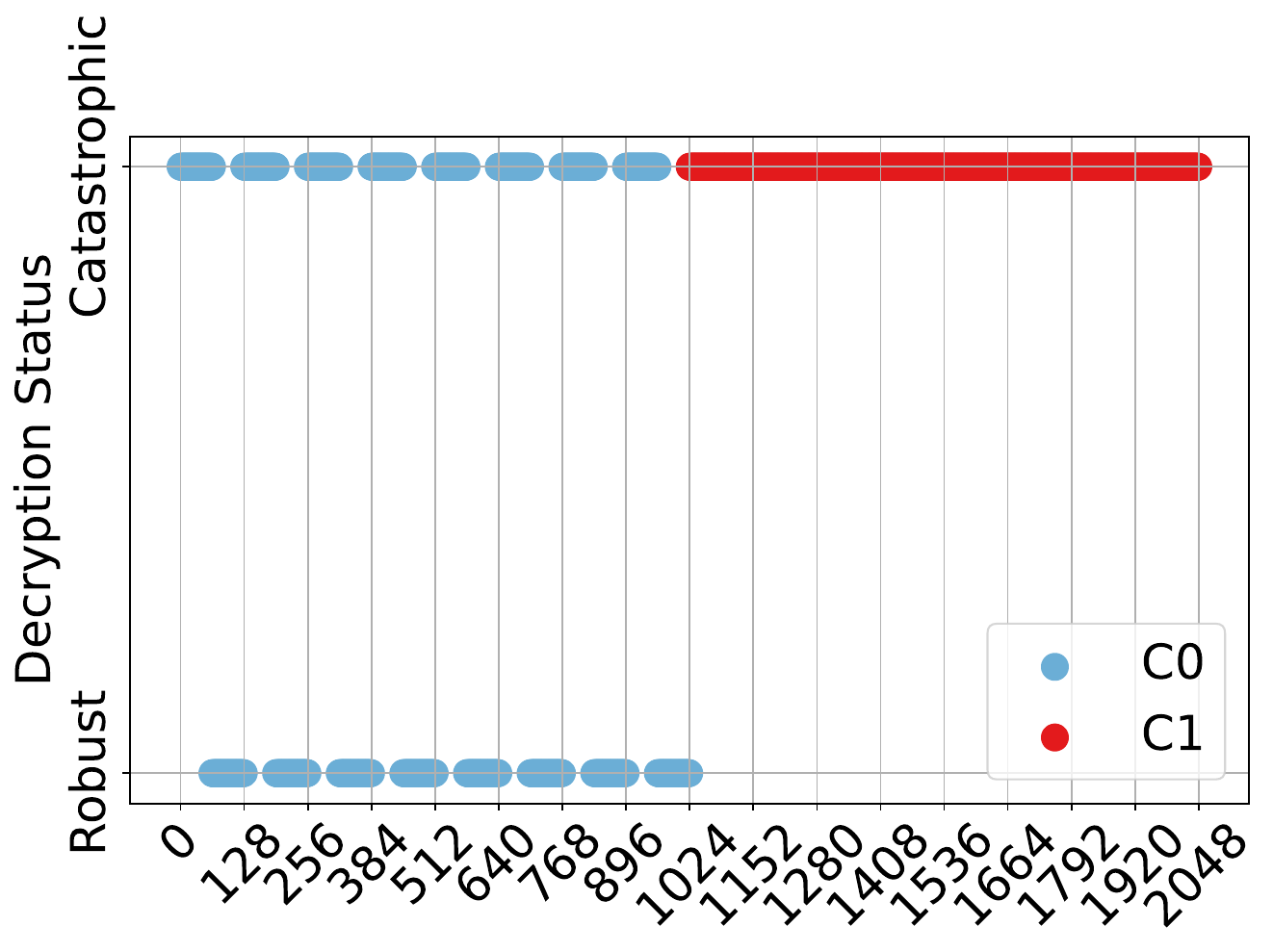}
    \caption{
        Bit flip during encryption on RNS-only mode with two limbs. The horizontal axis indicates the flipped bit position; the vertical axis indicates whether decoding succeeded (robust category) or failed (catastrophic category). Parameters: 60‑bit $q_0$, input range from \numrange{0}{256}, \(\Delta=2^{20}\), \(N=16\), and four encrypted elements.}
    \label{fig:EncryptNorm_openfhe_RNS}
\end{figure}

We focus on  the RNS-only configuration \textit{when the number of encrypted elements is reduced to half or less of the available encryptable elements}.
Figure~\ref{fig:EncryptNorm_openfhe_RNS} reports results for ring dimension \(N=16\) with only four encrypted elements, scaling factor \(\Delta=2^{20}\), and 8‐bit input values.
We observe that \emph{odd}‐indexed coefficients of \(c_0\) remain fully robust: all the input elements decrypt correctly.
However, any bit flip in an \emph{even}‐indexed coefficient, causes catastrophic failures.
The number of sensitive coefficients exactly matches the \textit{gap} analysis from Section~\ref{subsec:encGap}. Polynomial \(c_1\) exhibits no robust bits at all: every bit flip causes decrypt failure.

These results show that running the RNS-only mode leaving some unused elements can unveil robustness.
To ensure a secure-enough schemes, CKKS requires a ring dimension in the order of $2^{15}$, which leads to a payload of near $2^{14}$ encrytable elements.
In many practical uses, the amount of needed elements is significantly less, allowing a fault tolerance scheme by taking advantage of the unsed input elements.

\textbf{MNIST Example}.
To illustrate the practical impact of bit-flip resilience in CKKS, we consider a case using input images from the MNIST dataset~\cite{Deng2012}.
Each image consists of $28 \times 28 = 784$ grayscale pixels, where each pixel is encoded using integer values ranging from 0 to 255.

To encode 784 values using CKKS batching, we need a vector size that is a power of two and greater than or equal to 784; the smallest size is $2^{10} = 1024$.
Since CKKS requires a ring dimension $N$ that is twice the batch size, we need $N \geq 2^{11} = 2048$.
However, to comply with security requirements, we select a ring dimension of $N = 2^{15}$.
Only $2^{11}$ coefficients of the plaintext or ciphertext polynomials store actual image data, while the remaining $2^{15} - 2^{11}$ coefficients are unused.

\begin{figure}
    \centering
    \includegraphics[width=0.85\linewidth]{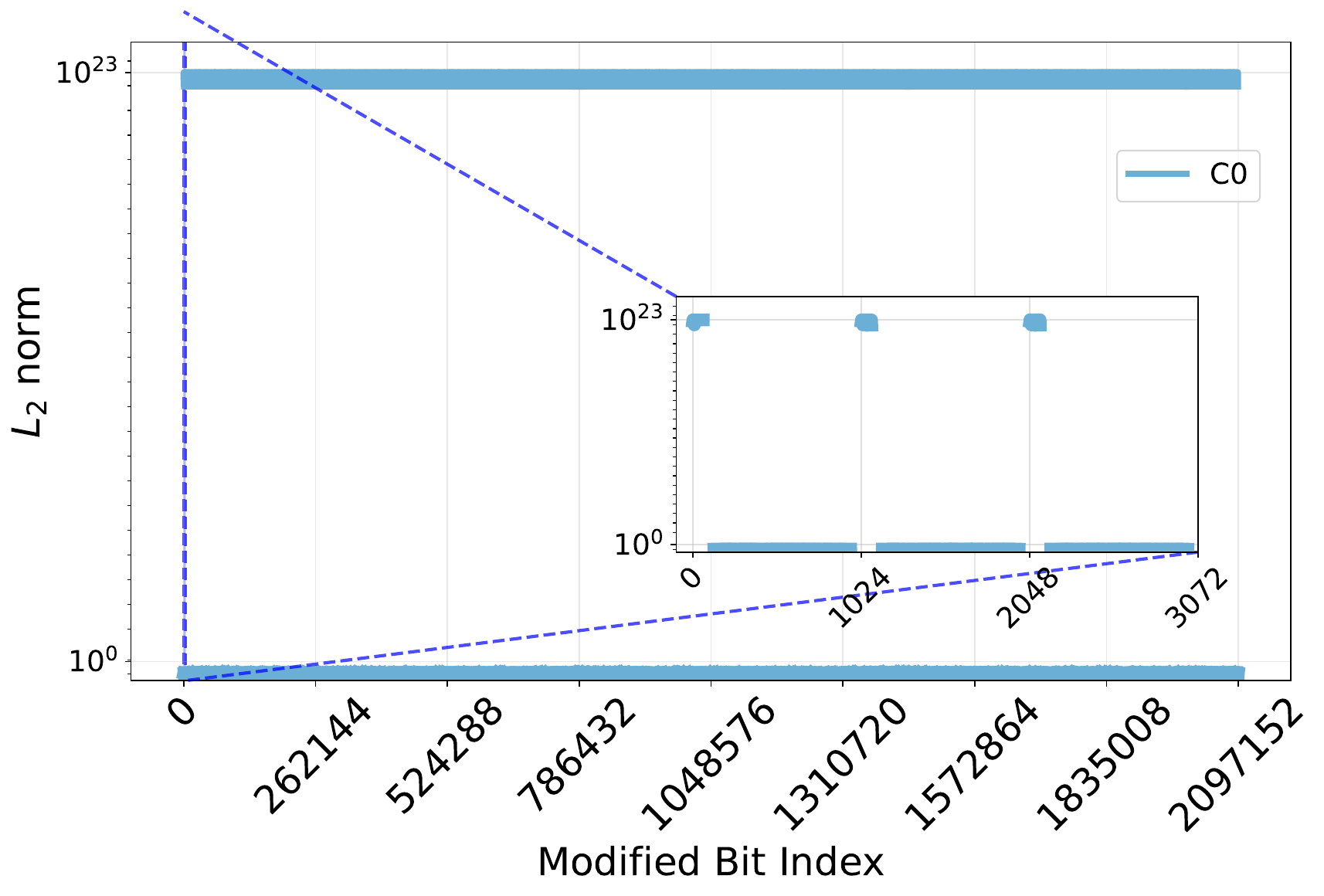}
    \caption{Bit flip during encryption on RNS-only mode with three limbs, using MNIST images as input. The horizontal axis indicates the flipped bit position; the vertical axis indicates the resulting L2 norm of the decoding error. Only $c_0$ is shown because $c_1$ is all undecryptable similarly as Fig.~\ref{fig:EncryptNorm_openfhe_RNS}. Parameters: 60-bit $q_0$,  $\Delta = 2^{20}$,  $N = 2^{15}$, and $2^{10}$ encrypted elements.}
    \label{fig:mnist}
\end{figure}

Figure~\ref{fig:mnist} shows results for $N=2^{15}$, and $2^{10}$ encrypted elements of MNIST images, $q_0$ modulus of 60 bits, scaling factor of $2^{20}$, and three-limb RNS.
As expected, flipping any bit in the $c_1$ polynomial consistently produces an undecryptable result, whereas in $c_0$ only about one in every sixteen coefficients is sensitive to bit-flips.

For this configuration, $\textit{gap} = \frac{N/2}{\text{needed elements}}=\frac{2^{15}/2}{2^{10}} = 16$.
Therefore, only $N/16 = 2048$ coefficients in the $c_0$ polynomial would be affected by a bit flip, while the rest remain unaffected.
In other words, over \SI{93.75}{\percent} of the encoded coefficients in the plaintext, or \SI{46.87}{\percent} of the coefficients in the ciphertex\footnote{All $c_1$ coefficients remain sensitive to bit errors.}, are inherently resilient to single-bit or multi-bit faults in this scenario.
However, this resiliency comes at a substantial computational cost.
As explained in Section~\ref{section:background}, NTT-based multiplication enables fast polynomial product computations with a complexity of $\mathcal{O}(N \log N)$, where $N$ is the ring dimension.
When NTT is disabled, the underlying polynomial multiplication reverts to a schoolbook approach with complexity $\mathcal{O}(N^2)$.
As a result, homomorphic operations become orders of magnitude slower when operating in coefficient space without NTT.

Although we do not perform direct runtime benchmarking in this study, prior works such as~\cite{Cheon2019} report performance slowdowns of approximately ten times when switching from the RNS+NTT-optimized scheme to the vanilla CKKS implementation.
In scenarios where RNS is used but NTT is disabled, the degradation is expected to be smaller, since RNS still reduces the cost of large-integer arithmetic by enabling parallel residue computations.
In practice, the slowdown depends on various factors, including the specific implementation, the chosen ring dimension, and the availability of hardware-level optimizations (e.g., SIMD instructions or parallel FFTs).
This highlights a fundamental trade-off between fault resilience and computational efficiency, which must be carefully balanced in real-world deployments.

\section{Conclusions}\label{section:conclusions}
This paper presents a comprehensive evaluation of CKKS's sensitivity to single‐bit flips, emphasizing the influence of RNS and NTT optimizations on error resilience.
While the vanilla CKKS implementation exhibits substantial inherent tolerance to individual bit‐flip faults, the optimized variants--with RNS or NTT enabled--show markedly increased vulnerability, potentially undermining the correctness of decrypted outputs.

We provide a high‑level comparison of the different CKKS configurations evaluated in this work. For each scenario, we highlight the key resilience pattern and any notable observations.
\paragraph*{Vanilla CKKS}
Least‐significant bits exhibit robustness; error grows exponentially for higher bits up to modulus $q_0$.
Coefficient \(N/2\) robust in HEAAN~1.0 but not in OpenFHE vanilla (due to Gaussian noise injection).
\paragraph*{Varying \(\Delta\)}
Larger \(\Delta\) shifts the onset of exponential error growth towards more significant bits.
\(\Delta \ge 2^{30}\) yields robust outcomes in over 80\% of trials.
\paragraph*{Effective Representation Range}
Increasing integer‐part range raises catastrophic frequency and reduces robust frequency.
Behavior mimics that of increasing \(\Delta\).
\paragraph*{Fault tolerance based on unused input elements}
Encoding fewer elements increases the number of robust coefficients in \(c_0\) or plaintext.
Not using NTT increases substantially the computational cost.

Our findings indicate that, despite its performance drawbacks, vanilla CKKS offers a robustness advantage by minimizing the impact of isolated bit errors compared to its optimized counterparts.
Furthermore, when employing large ring dimensions for security, encoding fewer elements (powers of two below the maximum) introduces redundancy that renders a subset of coefficients robust to bit-flips.
This \textit{gap}‐driven effect persists even in some optimized configurations, highlighting a practical mechanism for enhancing resilience at the cost of performance.

These results provide actionable guidance for balancing throughput, security, and error resilience when selecting CKKS parameters.
They also lay the groundwork for future research in the context of application-specific error tolerance--e.g., deep neural networks for image processing--where limited accuracy loss is acceptable, as well as evaluation in multi-bit error scenarios.

\section*{Acknowledgments}
We acknowledge the support from the open‐source homomorphic encryption community and the developers of the OpenFHE and HEAAN libraries.
We extend our gratitude to Dr.~Ahmad Al Badawi for his help setting up OpenFHE and clear explanations.
This work is partially supported by grants from Universidad de Buenos Aires (UBA) UBACYT 20620190100001BA, and CONICET PIP 11220220100027CO.

\bibliographystyle{IEEEtran}
\bibliography{journalsAbbr,fhe}

\begin{thebibliography}{10}
\providecommand{\url}[1]{#1}
\csname url@samestyle\endcsname
\providecommand{\newblock}{\relax}
\providecommand{\bibinfo}[2]{#2}
\providecommand{\BIBentrySTDinterwordspacing}{\spaceskip=0pt\relax}
\providecommand{\BIBentryALTinterwordstretchfactor}{4}
\providecommand{\BIBentryALTinterwordspacing}{\spaceskip=\fontdimen2\font plus
\BIBentryALTinterwordstretchfactor\fontdimen3\font minus
  \fontdimen4\font\relax}
\providecommand{\BIBforeignlanguage}[2]{{%
\expandafter\ifx\csname l@#1\endcsname\relax
\typeout{** WARNING: IEEEtran.bst: No hyphenation pattern has been}%
\typeout{** loaded for the language `#1'. Using the pattern for}%
\typeout{** the default language instead.}%
\else
\language=\csname l@#1\endcsname
\fi
#2}}
\providecommand{\BIBdecl}{\relax}
\BIBdecl

\bibitem{Cheon2017}
J.~H. Cheon, A.~Kim, M.~Kim, and Y.~Song, ``Homomorphic encryption for
  arithmetic of approximate numbers,'' in \emph{Proc of the Int Conf on the
  Theory and Applications of Cryptology and Information Security (ASIACRYPT)},
  T.~Takagi and T.~Peyrin, Eds., vol. 10624.\hskip 1em plus 0.5em minus
  0.4em\relax Cham: Springer International Publishing, 2017, pp. 409--437.

\bibitem{Rivest1978}
R.~L. Rivest, A.~Shamir, and L.~Adleman, ``A method for obtaining digital
  signatures and public-key cryptosystems,'' \emph{Commun ACM}, vol.~21, no.~2,
  p. 120–126, Feb. 1978.

\bibitem{Ma2022}
J.~Ma, S.-A. Naas, S.~Sigg, and X.~Lyu, ``Privacy-preserving federated learning
  based on multi-key homomorphic encryption,'' \emph{Int J Intell Syst},
  vol.~37, no.~9, pp. 5880--5901, 2022.

\bibitem{Zhang2023}
Y.~Zhang, Y.~Miao, X.~Li, L.~Wei, Z.~Liu, K.-K.~R. Choo, and R.~H. Deng,
  ``Efficient privacy-preserving federated learning with improved compressed
  sensing,'' \emph{IEEE T Ind Inf}, vol.~20, no.~3, pp. 3316--3326, 2024.

\bibitem{Zhang2024}
M.~Zhang, L.~Wang, X.~Zhang, Z.~Liu, Y.~Wang, and H.~Bao, ``Efficient
  clustering on encrypted data,'' in \emph{Applied Cryptography and Network
  Security}, C.~P{\"o}pper and L.~Batina, Eds.\hskip 1em plus 0.5em minus
  0.4em\relax Cham: Springer Nature Switzerland, 2024, pp. 213--236.

\bibitem{Cheon2019}
J.~H. Cheon, K.~Han, A.~Kim, M.~Kim, and Y.~Song, ``A full {RNS} variant of
  approximate homomorphic encryption,'' in \emph{Proc of the Int Conf on
  Selected Areas in Cryptography (SAC)}, C.~Cid and M.~J. Jacobson~Jr., Eds.,
  vol. 11349.\hskip 1em plus 0.5em minus 0.4em\relax Cham: Springer
  International Publishing, 2019, pp. 347--368.

\bibitem{Sabath2010}
F.~Sabath, \emph{Classification of Electromagnetic Effects at System
  Level}.\hskip 1em plus 0.5em minus 0.4em\relax New York, NY: Springer New
  York, NY, 2010, pp. 325--333.

\bibitem{Kolditz2014}
T.~Kolditz, T.~Kissinger, B.~Schlegel, D.~Habich, and W.~Lehner, ``Online bit
  flip detection for in-memory b-trees on unreliable hardware,'' in \emph{Proc
  of the Tenth Int Workshop on Data Management on New Hardware}, ser. DaMoN
  '14.\hskip 1em plus 0.5em minus 0.4em\relax New York, NY, USA: ACM, 2014, pp.
  1--9.

\bibitem{Mutlu2019rowhammer}
O.~Mutlu and J.~S. Kim, ``Rowhammer: A retrospective,'' \emph{IEEE T Comput Aid
  D}, vol.~39, no.~8, pp. 1555--1571, 2020.

\bibitem{Li2020}
K.-J. Li, Y.-Z. Xie, F.~Zhang, and Y.-H. Chen, ``Statistical inference of
  serial communication errors caused by repetitive electromagnetic
  disturbances,'' \emph{IEEE Electromagn Compat M}, vol.~62, no.~4, pp.
  1160--1168, 2020.

\bibitem{Li2020b}
Z.~Li, H.~Menon, D.~Maljovec, Y.~Livnat, S.~Liu, K.~Mohror, P.-T. Bremer, and
  V.~Pascucci, ``Spotsdc: Revealing the silent data corruption propagation in
  high-performance computing systems,'' \emph{IEEE T Visual Comput Graph},
  vol.~27, no.~10, pp. 3938--3952, 2020.

\bibitem{Hoeffgen2020}
S.~K. H{\"o}effgen, S.~Metzger, and M.~Steffens, ``Investigating the effects of
  cosmic rays on space electronics,'' \emph{Front Phys}, vol.~8, p. 318, 2020.

\bibitem{Dixit2021}
H.~Dixit, ``Silent data corruptions at scale,'' in \emph{Proc of the Int Symp
  on On-Line Testing and Robust System Design (IOLTS)}.\hskip 1em plus 0.5em
  minus 0.4em\relax IEEE, 2023, pp. 1--2.

\bibitem{Cheon2018}
J.~H. Cheon, K.~Han, A.~Kim, M.~Kim, and Y.~Song, ``Bootstrapping for
  approximate homomorphic encryption,'' in \emph{Advances in Cryptology --
  EUROCRYPT 2018}, J.~B. Nielsen and V.~Rijmen, Eds.\hskip 1em plus 0.5em minus
  0.4em\relax Cham: Springer International Publishing, 2018, pp. 360--384.

\bibitem{Bernstein2007}
D.~J. Bernstein, ``The tangent {FFT},'' in \emph{Proc of the Int Symp of
  Applied Algebra, Algebraic Algorithms and Error-Correcting Codes (AAECC)},
  ser. Lecture Notes in Computer Science, S.~Bozta{\c{s}} and H.-F.~F. Lu,
  Eds., vol. 4851.\hskip 1em plus 0.5em minus 0.4em\relax Berlin, Heidelberg:
  Springer, 2007, pp. 291--300.

\bibitem{Bernstein2008}
------, \emph{\BIBforeignlanguage{English}{Fast multiplication and its
  applications}}, ser. Mathematical Sciences Research Institute
  Publications.\hskip 1em plus 0.5em minus 0.4em\relax United Kingdom:
  Cambridge University Press, 2008, pp. 325--384.

\bibitem{Lyubashevsky2013b}
V.~Lyubashevsky, C.~Peikert, and O.~Regev, ``A toolkit for ring-lwe
  cryptography,'' in \emph{Proc of the Annual Int Conf on the theory and
  applications of cryptographic techniques (EUROCRYPT)}, vol. 7881.\hskip 1em
  plus 0.5em minus 0.4em\relax Berlin, Heidelberg: Springer, 2013, pp. 35--54.

\bibitem{Costache2024}
A.~Costache, B.~R. Curtis, E.~Hales, S.~Murphy, T.~Ogilvie, and R.~Player, ``On
  the precision loss in approximate homomorphic encryption,'' in \emph{Proc of
  the Int Conf on Selected Areas in Cryptography (SAC)}.\hskip 1em plus 0.5em
  minus 0.4em\relax Berlin, Heidelberg: Springer-Verlag, 2024, p. 325–345.

\bibitem{VanderLeest2010}
V.~van~der Leest, G.-J. Schrijen, H.~Handschuh, and P.~Tuyls, ``Hardware
  intrinsic security from d flip-flops,'' in \emph{Proc of the Fifth ACM
  Workshop on Scalable Trusted Computing}, ser. STC '10.\hskip 1em plus 0.5em
  minus 0.4em\relax New York, NY, USA: ACM, Oct. 2010, p. 53–62.

\bibitem{Ziegler1979}
J.~F. Ziegler and W.~A. Lanford, ``Effect of cosmic rays on computer
  memories,'' \emph{Science}, vol. 206, no. 4420, pp. 776--788, 1979.

\bibitem{Mohan2016}
P.~V.~A. Mohan, \emph{Residue Number Systems: Theory and Applications}.\hskip
  1em plus 0.5em minus 0.4em\relax Birkh\"auser Chama, 2016.

\bibitem{AlBadawi2022}
\BIBentryALTinterwordspacing
A.~A. Badawi, A.~Alexandru, J.~Bates, F.~Bergamaschi, D.~B. Cousins,
  S.~Erabelli, N.~Genise, S.~Halevi, H.~Hunt, A.~Kim, Y.~Lee, Z.~Liu,
  D.~Micciancio, C.~Pascoe, Y.~Polyakov, I.~Quah, S.~R.V., K.~Rohloff,
  J.~Saylor, D.~Suponitsky, M.~Triplett, V.~Vaikuntanathan, and V.~Zucca,
  ``{OpenFHE}: Open-source fully homomorphic encryption library,'' Cryptology
  {ePrint} Archive, Paper 2022/915, 2022. [Online]. Available:
  \url{https://eprint.iacr.org/2022/915}
\BIBentrySTDinterwordspacing

\bibitem{Deng2012}
L.~Deng, ``The {MNIST} database of handwritten digit images for machine
  learning research,'' \emph{IEEE Signal Proc Mag}, vol.~29, no.~6, pp.
  141--142, 2012.

\end{thebibliography}

\begin{IEEEbiography}[{\includegraphics[width=1in,height=1in,clip]{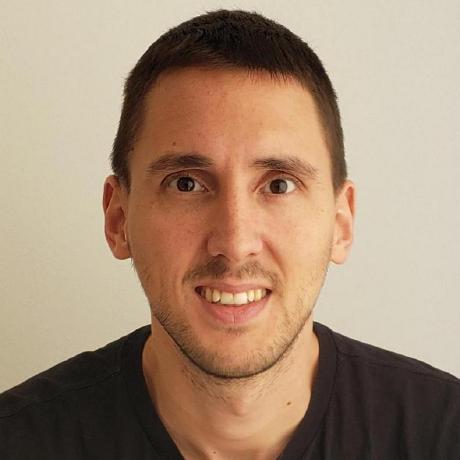}}]{Matias Mazzanti}
received  his master degree on Physics from Universidad de Buenos Aires in 2021. He is currently pursuing his Ph.D. in computer science at Universidad de Buenos Aires.
\end{IEEEbiography}
\begin{IEEEbiography}[{\includegraphics[width=1in,height=1in,clip]{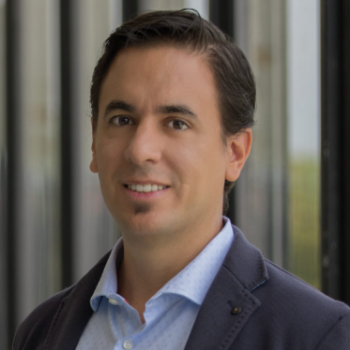}}]{Augusto Vega}
is a Senior Research Staff Member at IBM Thomas J. Watson Research Center, Yorktown Heights, NY, USA, involved in research and development work in the areas of heterogeneous systems and edge AI computing.  Vega received a Ph.D. degree from the Polytechnic University of Catalonia, Barcelona, Spain, in 2013.
\end{IEEEbiography}
\begin{IEEEbiography}[{\includegraphics[width=1in,height=1in,clip]{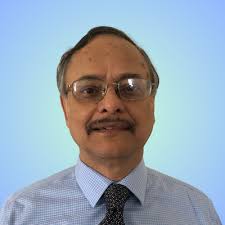}}]{Pradip Bose}
is a distinguished research staff member and the manager at IBM Thomas J. Watson Research Center, IBM, Yorktown Heights, NY 10598 USA. He is a Fellow of IEEE.
\end{IEEEbiography}
\begin{IEEEbiography}[{\includegraphics[width=1in,height=1in,clip]{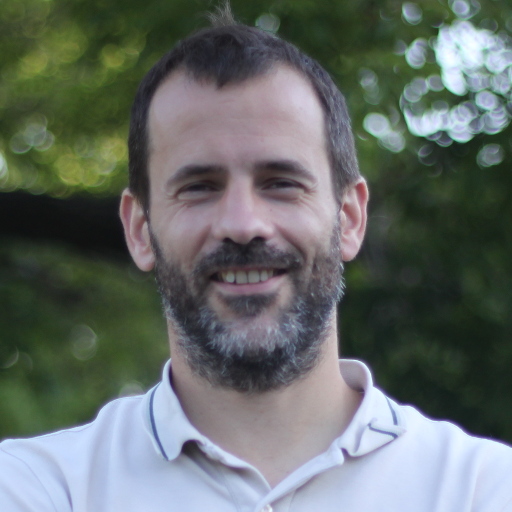}}]{Esteban Mocskos}
is a full-time professor at UBA and researcher at CSC-CONICET. He received his Ph.D. in Computer Science from UBA in 2008. His research interests are in the areas of distributed systems, computer networks and protocols, parallel programming and applications.
\end{IEEEbiography}

\end{document}